\documentclass[twoside,twocolumn,9pt]{article}

\usepackage{extsizes}
\usepackage[super,sort&compress,comma]{natbib}
\usepackage[version=3]{mhchem}
\usepackage[left=1.5cm, right=1.5cm, top=1.785cm, bottom=2.0cm]{geometry}
\usepackage{balance}
\usepackage{mathptmx}
\usepackage{sectsty}
\usepackage{graphicx}
\usepackage{lastpage}
\usepackage[format=plain,justification=justified,singlelinecheck=false,font={stretch=1.125,small,sf},labelfont=bf,labelsep=space]{caption}
\usepackage{float}
\usepackage{fancyhdr}
\usepackage{fnpos}
\usepackage[english]{babel}
\addto{\captionsenglish}{%
  \renewcommand{\refname}{Notes and references}
}
\usepackage{array}
\usepackage{droidsans}
\usepackage{charter}
\usepackage[usenames,dvipsnames]{xcolor}
\usepackage{setspace}
\usepackage[compact]{titlesec}
\usepackage{hyperref}
\definecolor{cream}{RGB}{222,217,201}
\usepackage[T1]{fontenc}

\usepackage{amsmath}
\usepackage{amssymb}
\usepackage{indentfirst}
\usepackage{microtype}
\usepackage{etoolbox}
\apptocmd{\sloppy}{\hbadness 10000\relax}{}{}

\usepackage{doi}
\makeatother
\usepackage{csquotes}
\usepackage{siunitx}
\DeclareSIUnit\C{C}
\DeclareSIPrefix\micro{\text{\textmu}}{-3}
\usepackage{tabularx}
\usepackage{multirow}
\usepackage{prettyref}
\newrefformat{sec}{Section \nameref{#1}}
\newrefformat{subsec}{Subsection \nameref{#1}}
\newrefformat{eq}{\cref{#1}}
\newrefformat{fig}{Figure \ref{#1}}
\newrefformat{tab}{Table \ref{#1}}
\usepackage{todonotes}
\usepackage{tcolorbox}

\begin{document}

\pagestyle{fancy}
\thispagestyle{plain}
\fancypagestyle{plain}{
\renewcommand{\headrulewidth}{0pt}
}

\makeFNbottom
\makeatletter
\renewcommand\LARGE{\@setfontsize\LARGE{15pt}{17}}
\renewcommand\Large{\@setfontsize\Large{12pt}{14}}
\renewcommand\large{\@setfontsize\large{10pt}{12}}
\renewcommand\footnotesize{\@setfontsize\footnotesize{7pt}{10}}
\makeatother

\renewcommand{\thefootnote}{\fnsymbol{footnote}}
\renewcommand\footnoterule{\vspace*{1pt}%
\color{cream}\hrule width 3.5in height 0.4pt \color{black}\vspace*{5pt}} 
\setcounter{secnumdepth}{5}

\makeatletter 
\renewcommand\@biblabel[1]{#1}            
\renewcommand\@makefntext[1]%
{\noindent\makebox[0pt][r]{\@thefnmark\,}#1}
\makeatother 
\renewcommand{\figurename}{\small{Fig.}~}
\sectionfont{\sffamily\Large}
\subsectionfont{\normalsize}
\subsubsectionfont{\bf}
\setstretch{1.125} 
\setlength{\skip\footins}{0.8cm}
\setlength{\footnotesep}{0.25cm}
\setlength{\jot}{10pt}
\titlespacing*{\section}{0pt}{4pt}{4pt}
\titlespacing*{\subsection}{0pt}{15pt}{1pt}

\fancyfoot{}
\fancyfoot[LO,RE]{\vspace{-7.1pt}\includegraphics[height=9pt]{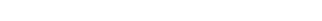}}
\fancyfoot[CO]{\vspace{-7.1pt}\hspace{13.2cm}\includegraphics{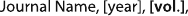}}
\fancyfoot[CE]{\vspace{-7.2pt}\hspace{-14.2cm}\includegraphics{head_foot/RF}}
\fancyfoot[RO]{\footnotesize{\sffamily{1--\pageref{LastPage} ~\textbar  \hspace{2pt}\thepage}}}
\fancyfoot[LE]{\footnotesize{\sffamily{\thepage~\textbar\hspace{3.45cm} 1--\pageref{LastPage}}}}
\fancyhead{}
\renewcommand{\headrulewidth}{0pt} 
\renewcommand{\footrulewidth}{0pt}
\setlength{\arrayrulewidth}{1pt}
\setlength{\columnsep}{6.5mm}
\setlength\bibsep{1pt}

\makeatletter 
\newlength{\figrulesep} 
\setlength{\figrulesep}{0.5\textfloatsep} 

\newcommand{\topfigrule}{\vspace*{-1pt}%
\noindent{\color{cream}\rule[-\figrulesep]{\columnwidth}{1.5pt}} }

\newcommand{\botfigrule}{\vspace*{-2pt}%
\noindent{\color{cream}\rule[\figrulesep]{\columnwidth}{1.5pt}} }

\newcommand{\dblfigrule}{\vspace*{-1pt}%
\noindent{\color{cream}\rule[-\figrulesep]{\textwidth}{1.5pt}} }

\makeatother

\twocolumn[
  \begin{@twocolumnfalse}
{\includegraphics[height=30pt]{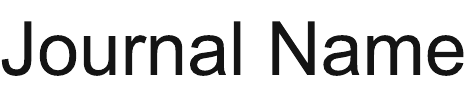}\hfill\raisebox{0pt}[0pt][0pt]{\includegraphics[height=55pt]{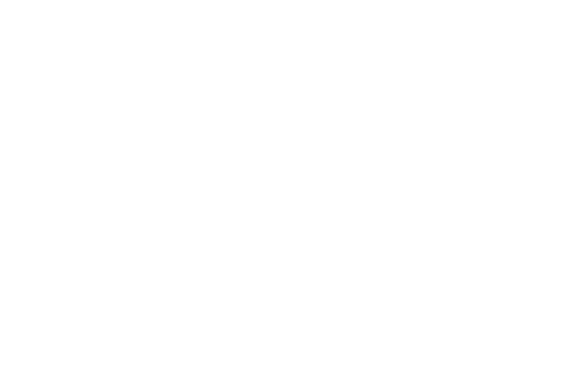}}\\[1ex]
\includegraphics[width=18.5cm]{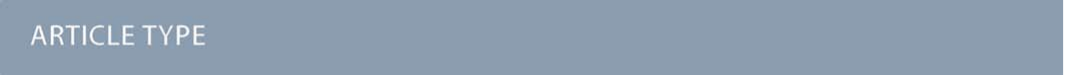}}\par
\vspace{1em}
\sffamily
\begin{tabular}{m{4.5cm} p{13.5cm} }

\includegraphics{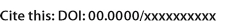} & \noindent\LARGE{\textbf{Workflows and Principles for Collaboration and Communication in Battery Research$^\dag$}} \\
\vspace{0.3cm} & \vspace{0.3cm} \\

 & \noindent\large{Yannick Kuhn,\textit{$^{a}$} Bhawna Rana,\textit{$^{b}$} Micha Philipp,\textit{$^{a}$} Christina Schmitt,\textit{$^{b}$} Roberto Scipioni,\textit{$^{c}$} Eibar Flores,\textit{$^{d}$} Dennis Kopljar,\textit{$^{b}$} Simon Clark,\textit{$^{d}$} Arnulf Latz,\textit{$^{a}$} and Birger Horstmann$^{\ast}$\textit{$^{a}$}} \\

\includegraphics{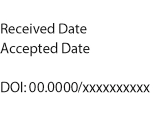} & \noindent\normalsize{Interdisciplinary collaboration in battery science is required for rapid evaluation of better compositions and materials. However, diverging domain vocabulary and non-compatible experimental results slow down cooperation. We critically assess the current state-of-the-art and develop a structured data management and interpretation system to make data curation sustainable. The techniques we utilize comprise ontologies to give a structure to knowledge, database systems tenable to the FAIR principles, and software engineering to break down data processing into verifiable steps. To demonstrate our approach, we study the applicability of the Galvanostatic Intermittent Titration Technique on various electrodes. Our work is a building block in making automated material science scale beyond individual laboratories to a worldwide connected search for better battery materials.} \\

\end{tabular}

 \end{@twocolumnfalse} \vspace{0.6cm}

  ]

\renewcommand*\rmdefault{bch}\normalfont\upshape
\rmfamily
\section*{}
\vspace{-1cm}


\footnotetext{\textit{$^{a}$~German Aerospace Center, Wilhelm-Runge-Stra{\ss}e 10, 89081 Ulm, Germany. E-mail: birger.horstmann@dlr.de}}
\footnotetext{\textit{$^{b}$~Deutsches Zentrum für Luft- und Raumfahrt (DLR), Pfaffenwaldring 38-40, 70569 Stuttgart, Germany.}}
\footnotetext{\textit{$^{c}$~SINTEF Energy Research, Department of Thermal Energy, Kolbjørn Hejes vei 1 A, 7034 Trondheim, Norway.}}
\footnotetext{\textit{$^{d}$~SINTEF Industry, Department of Sustainable Energy Technology, 7491 Trondheim, Norway.}}

\footnotetext{\dag~Supplementary Information available: [details of any supplementary information available should be included here]. See DOI: 00.0000/00000000.}




\section{Introduction}

Li-ion batteries are widely deployed for transportation, grid stabilization, and power tools.
These applications require specialized batteries, e.g., with long service life, or performance under extreme conditions.
Enhancing the usable lifespan and power density of future batteries will greatly aid in their successful deployment. \par
Discovering new battery compositions requires a comprehensive understanding of the physicochemical processes taking place within them. Building this understanding starting from a candidate battery material requires thorough collaboration across a wide variety of disciplines: chemists to formulate the material, physicists to develop the experimental machinery, experimental electrochemists to perform and interpret the experiments, and theoretical electrochemists to find patterns in the data. Due to the many mechanisms that happen in a battery during this chain of events \cite{Xu2024}, interdisciplinary communication between all scientists involved in battery characterization is needed. \par
However, the exchange between theoretical and experimental electrochemistry has always been challenging due to a divergence in discipline-specific language and domain knowledge. This is a direct consequence of the necessarily different challenges in battery characterization and the diverse backgrounds. On the one hand, model-based interpretation of experiments introduces bias by assigning processes to measurement features. However, the underlying assumptions may not be transparent. On the other hand, most experimental procedures have to be repeated multiple times to overcome challenges around accuracy and reproducibility. Yet, the concise publication of a representative single dataset may not make that transparent. \par
To bring the community forward and better align theoretical and experimental efforts, workflows and methods need to be established, with which we can produce high-quality data in an up-scalable manner while increasing its compliance with the FAIR principles \cite{Wilkinson2016}: Findable, Accessible, Interoperable, and Reusable. \par
Current efforts to achieve this task span many disciplines and problems. Describing the data such that it can be collected and analyzed across instutions requires common requirements and a formalized language, e.g., ontologies. \cite{Mistry2021a, Dechent2021, Clark2022} Maximizing the throughput of any one institution is aided by automation and digitalization of the experiments. \cite{Dave2020, Lewis-Douglas2020, Vogler2024, Stier2024} Interpreting the rich amount of data requires a close integration with advanced Machine Learning techniques. \cite{Houchins2020, Sanin2025} \par
This paper demonstrates a semi-autonomous, FAIR-compliant workflow to create reusable measurement data and use it to parameterize electrochemical battery models. We show how it allows us to identify and resolve mismatching biases in data interpretation.
First, we detail the theory behind the battery models we use, the measurement techniques we employ, the FAIR principles, and the algorithms we use for parameterization.
Second, we present a case study of how we elucidated a commonly occurring mismatch between active material diffusivities, depending on the measurement technique or theoretical model treatment.
Third, we report on our findings on enhancing the collaboration between experimentalists and theoreticians.
Finally, we conclude with a summary of our findings. \par

\section{Methods} \label{sec:Methods}

\subsection{Ensuring FAIR Workflows}

We present the methods we investigated to apply the FAIR principles in practice for battery measurements.
We implement the four methods to improve compliance with the FAIR principles: data annotation with ontologies, data publication in open repositories, automated data processing workflows, and data review. \par
Findability requires structured metadata that make datasets searchable and discoverable. We achieve this by creating metadata based on key-value pairs. The structure and terms used in the metadata are taken from the Battery Interface Ontology (BattINFO). \cite{Dechent2024}
BattINFO is a domain ontology that expresses knowledge about batteries using a formal and machine-readable vocabulary. 
It is an extension of the Elementary Multiperspective Materials Ontology (EMMO). 
This allows metadata annotated with BattINFO terms to be understood within the broader scope of physics and materials science and enables interoperability with other datasets annotated with EMMO. Furthermore, annotating datasets with semantic vocabularies and linking to other datasets adheres to World Wide Web Consortium (W3C) recommendations for publishing linked data on the Web, which extends the findability for web-based queries.
The structured and semantically annotated metadata is serialized as JSON-LD files and stored alongside the data using the database software Kadi4Mat \cite{Brandt2021}. \par
Accessibility ensures that datasets, once discovered, can be retrieved.
Open data repositories act as publishers of datasets, providing long-term storage, persistent identifiers, and version control.
Using third-party repositories such as Zenodo, operated by CERN, suitable datasets can be made available with central access venues.
Kadi4Mat simplifies this process with a three-click integration, preserving metadata and data files for immediate access and citation. \par
Interoperability allows datasets and workflows to be reused in new contexts.
We achieve this by adhering to standardized formats with ontology-annotated data models and breaking the data processing pipeline into modular workflows. 
These workflows define clear inputs, processes, and outputs, enabling automation and machine-readability.
Structuring data to support automation incidentally makes the process much more transparent and requires machine-readable (intermediate) results.
Kadi4Mat provides an infrastructure to keep the data and the workflows acting on it in one. \par
Reproducibility ensures that others can understand, verify, and adapt the developed workflows.
External reviews help validate data pipelines and uncover any missing documentation. 
For reusability, we incorporate a checklist-based review to confirm that datasets meet legal and practical requirements, such as licensing and accessibility. \par
Reusability ensures that datasets can be reused in future research, both legally and practically. To enable this, we provide datasets under a permissive Creative Commons license, which allows others to access, share, and adapt the data as long as proper credit is given. This licensing approach maximizes the potential for collaboration, innovation, and integration of our work into broader research efforts.

\subsection{Electrochemical Battery Parameterization} \label{subsec:ElectrochemicalTheory}

Our goal is to obtain material parameters that, when plugged into predictive models of the cell state, will give the results that we would observe in validation measurements.
On the cell level without access to microstructure imaging data, our most accurate dynamic model is the Doyle-Fuller-Newman (DFN) model \cite{Doyle1993}. For a thermodynamically consistent derivation and treatment of the general class of models that the DFN belongs to, we refer to Latz et al. \cite{Latz2011}.
Limitations of the DFN appear when considering the complex microstructures that arise in thick electrodes or novel materials \cite{Traskunov2021a}, as well as previously negligible effects in novel electrolytes \cite{Schammer2021}. To account for microstructure effects in the context of the DFN, please refer to Traskunov et al. \cite{Traskunov2021b}. Recent research showed that 3D microstructure models do not offer higher short-term voltage prediction accuracy than the DFN for commercial-like batteries \cite{Tredenick2023}. 3D microstructure models do, however, offer higher predictive capability for cell degradation \cite{Squires2024}. Hence, we do not consider microstructure effects here. \par
 \par
We demonstrate the challenges in systematically treating complex and diverse characterization data, starting with inverse modelling of the DFN as \textquote{most accurate method} and work our way down via simplifications of the DFN to direct parameter extraction from graphs.
The single particle model with electrolyte (SPMe) \cite{Marquis2019} is the linearized version of the DFN concerning electrolyte dynamics. Effectively, it approximates the DFN with only one representative particle per electrode, while resolving the electrolyte dynamics spatially. \par
The single particle model (SPM) \cite{Marquis2019} is the constant term of the DFN concerning electrolyte dynamics. It may be considered a DFN that neglects the electrolyte dynamics and treats the electrodes as one representative particle each.
We use PyBaMM \cite{Sulzer2021} to simulate these models. Marquis et al. \cite{Marquis2019} documented their equations and distinguishable parameter groupings. \par

\subsection{Measuring and Interpreting GITT Battery Response}

We now introduce the experimental methods and their interpretations we will consider. The Galvanostatic Intermittent Titration Technique (GITT) was introduced in 1977 to study molecule transport phenomena in electrochemistry. \cite{Weppner1977} With GITT, the battery experiences a short constant-current pulse, followed by a sufficiently long rest period.  GITT is used most commonly for the determination of diffusion coefficients. Please refer to the SI Subsection 2.4 for the formula and its modernization \cite{Chien2021, Kang2021}. Later in the paper, we will only use the square-root slope of the voltage signal shortly after current changes, abbreviated as \(\gamma := \partial U / \partial \sqrt t\). \par
Applying inverse modelling to GITT can utilize the measurement more comprehensively \cite{Kuhn2022}.
Escalante et al. \cite{Escalante2021} have already discussed the differences that can arise due to the model choice, in the case of the SPM vs. the SPMe. We will elaborate on that by additionally including the DFN. \par
GITT also yields the most accurate measurement of the Open-Circuit Potential (OCP) at any one State-of-Charge (SOC), i.e., the degree of lithiation between the maximally delithiated and maximally lithiated states. Measurement of voltage at a low current, typically cycling the battery in \qty{50}{\hour} or more (quasi-OCP), gives many SOC points but mixes static and dynamic parts and flattens features in the OCP curve. With GITT, the cell gets cycled with short constant-current pulses only changing the SOC by a small percentage value. Longer rest phases in-between let the voltage signal exponentially decay close to the OCP, and we take the exponential asymptote as the OCP. Hence, compared to quasi-OCP, the SOC resolution has to be lower, but each measurement is more accurate. One can alleviate this a bit by shortening the rest phases between GITT pulses and recovering their terminal voltage from exponential extrapolation \cite{Kuhn2022}. In any case, a reliable rest phase must have the material exhibit only one mode of exponential relaxation at its end. A plot of voltage over logarithmic time easily verifies this for graphite and NMC for rest phases as short as \qty{15}{\minute}; long-term hysteresis can require rest phases as long as weeks for other materials like silicon \cite{Köbbing2024, Wycisk2024}. \par
The algorithm we use to fit simulation models to data is called Expectation Propagation with Bayesian Optimization for Likelihood-Free Inference (EP-BOLFI). \cite{Kuhn2022}
It can tackle the high nonlinearity, i.e., the complexity of our models, while not only managing but also incorporating uncertainties in data, model, and model parameters into the fits.

\section{Experimental} \label{sec:Experimental}

\subsection{Cell composition}

We conduct our experiments on the INR18650-MJ1, a cylindrical \qty{3500}{\milli\ampere\hour} Li-ion battery cell in the 18650 format manufactured by LG Chem. With its high cycle life, an energy density of \qty{710}{\watt\hour\per\litre}, and a specific energy of \qty{260}{\watt\hour\per\kilogram}, measured at reference current \qty{0.2}{C}, it is often employed for high energy applications. The positive electrode active material is a high-nickel NMC-840511 (\ce{LiNi_{0.84}Mn_{0.05}Co_{0.11}O_2}) positive electrode, based on our measurements using inductively coupled plasma atomic emission spectroscopy (ICP-OES) performed on a Varian Vista-MPX. The element ratios are consistent with scanning electron microscopy (SEM) from a Zeiss Gemini Ultra plus and energy-dispersive X-ray spectroscopy (EDX) from a Bruker XFlash detector 5010, which report the average ratios N:M:C 83:5:12. Both ratio results are consistent with a report by Li et al. \cite{Li2019a}, stating N:M:C 82:6:11 from ICP and EDX. The negative electrode active material is a graphite/silicon oxide composite, as confirmed via SEM-EDX. The ratio of graphite to silicon oxide is determined from Micro Computer Tomography (µCT), as 96.5 volume-\% graphite and 3.5 volume-\% silicon oxide. Electrolyte harvested from the cell was measured via gas chromatography-mass spectrometry (GC-MS) by Sturm et al. \cite{Sturm2019}, revealing it to be \qty{1}{\mole\per\litre} \ce{LiPF_6} in a solvent based on EC:EMC:DMC with 1:1:1 volume ratios. The transport parameters for this electrolyte are well-documented and, therefore, taken from the literature. \cite{Schmalstieg2018} \par

\subsection{Cell disassembly}

To study materials and perform experiments on the electrode level, the cell is disassembled and the electrodes extracted. To this end, the cell is discharged to the discharge cut-off voltage of \qty{2.5}{\volt} at \qty{0.1}{\ampere} (\(C/50\)), transferred to an argon-filled glove box, and opened with a pipe cutter. After the cell is dismantled, positive and negative electrodes are extracted and carefully separated from the separator to avoid cross-contamination. For all measurements, the electrode and separator samples were washed twice for one minute with Dimethyl Carbonate (DMC) and dried in the glove box. For Electrochemical Impedance Spectroscopy (EIS) to measure tortuosities, to remove any residual salts, the electrodes were immersed in DMC overnight and left to dry for 30 minutes before re-assembly. For electrochemical experiments, the coating on one side of the double-sided electrodes must be removed from the current collector foil. The positive electrode coating is removed with N-Methyl-2-Pyrrolidone (NMP). In contrast, the coating of the negative electrode is removed outside of the glove box with deionized water, as a water-soluble binder is commonly used there. Images of the jelly roll removed from the cell can and the subsequent removal of the coatings are depicted in the SI Figure 1. The process of dismantling the commercial cell and preparing the components for different types of measurements is described in more detail in Schmitt et al. \cite{Schmitt2023}, together with a comprehensive description and assessment of various techniques for parameter identification. \par

\subsection{Cell geometry and microstructure}

Coating thicknesses are determined with the same SEM setup as earlier, a Zeiss Gemini Ultra plus with EDX from a Bruker XFlash detector 5010. The positive electrode has a \qty{73}{\micro\metre} coating thickness with \qty{19812}{\mole\per\cubic\metre} maximum lithium concentration according to SEM images of the electrode cross-section. As for the negative electrode, SEM images give a \qty{87}{\micro\metre} coating thickness with joint \qty{29254}{\mole\per\cubic\metre} maximum lithium concentration. The thicknesses are averaged over several SEM images to accommodate for local variations. Exemplary SEM images are shown in the SI Figure 3 to get an impression of particle morphology. The separator is a ceramic-coated polymer with a thickness of approximately \qty{12}{\micro\metre}. The images of the electrode surface reveal that the active materials on the negative and positive electrodes are in flake and spherical shape, respectively. Parameters describing the microstructure of the electrode are quantitatively assessed by analyzing the 3D reconstruction of the porous structure obtained by Focused Ion Beam nanotomography (FIB-nt). For that purpose, the microstructural data of the MJ1 cell provided in Heenan et al. \cite{Heenan2020a} is analyzed. For the reconstruction of the raw data, the 3D stack of images is segmented with the program ImageJ (from NIH). Due to non-uniform illumination, setting a single threshold for all micrographs is not feasible. Therefore, a Sauvola algorithm \cite{Sauvola2000} is used to perform local thresholding of the data. The Sauvola algorithm works by dividing the input image into square windows and setting thresholds for each based on the mean and standard deviation of the pixel intensities.

\begin{figure}[!ht]
    \centering
    \includegraphics[width=\columnwidth]{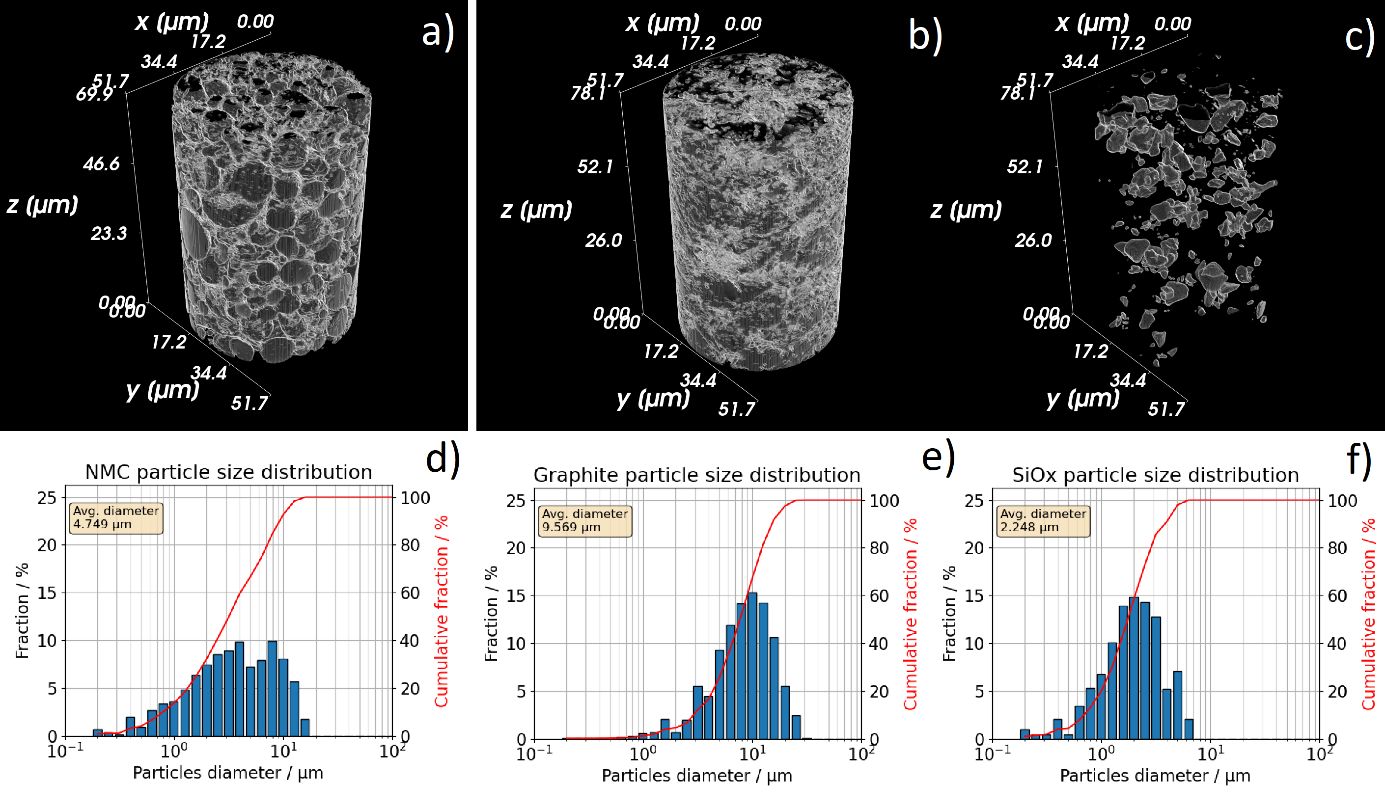}
    \caption{Visualization of the 3D reconstructions of the (a) NMC active phase in the positive electrode and of the (b) graphite and (c) Silicon-Oxide phases in the negative electrode. Below those are the calculated particle size distributions for the (d) NMC, (e) graphite, and (f) Silicon phases.}
    \label{fig:3D_reconstruction}
\end{figure}

\prettyref{fig:3D_reconstruction} shows the visualizations of the 3D reconstructions of the analyzed data. The data visualization uses Mayavi, a Python-based data visualization library. The Particle Size Distributions (PSD) of the various phases are calculated with MATLAB and TauFactor \cite{Cooper2016}, using a method introduced by Münch et al. \cite{Munch2008}. For a comprehensive microstructural analysis, we refer to Heenan et al. \cite{Heenan2020a}. The porosities of the electrode and separators are measured by mercury porosimetry \cite{Carniglia1986}, using the Pascal 140 + 240 system by Thermo Scientific, resulting in \(0.26\), \(0.38\), and \(0.23\), for the negative electrode, separator, and positive electrode, respectively. A maximum pressure of \qty{200}{\mega\pascal} is applied to the evacuated samples. \par

\begin{table}[ht]
    \small
    \caption{Path-length tortuosity \(\tau^2\), MacMullin number \(N_M\), and Bruggeman exponent \(\beta\) of different components of the MJ1 cell determined by EIS}
    \label{tab:geometry}
    \begin{tabular*}{\columnwidth}{@{\extracolsep{\fill}}llll}
    \hline
    Component & \(\tau^2\) & \(N_M\) & \(\beta\) \\
    \hline
    Negative el. & \(4.6\pm0.9\) & \(17.5\pm3.5\) & \(1.1\pm0.2\) \\
    Separator & \(4.64\pm0.05\) & \(12.2\pm0.1\) & \(1.59\pm0.01\) \\
    Positive el. & \(2.5\pm0.1\) & \(11.0\pm0.6\) & \(0.6\pm0.0\) \\
    \hline
    \end{tabular*}
\end{table}

\subsection{Electrochemical measurements}

For all electrochemical measurements, electrodes with \qty{18}{\milli\metre} diameter are punched out and assembled in ECC-PAT-Core-Cells (EL-CELL) in three-electrode configuration, if not otherwise mentioned. Setups relating to GITT also refer to the measurement where GITT and EIS were performed intermittently. The resulting capacity is \qty{12}{\milli\ampere\hour}, as estimated from an OCP model fit \cite{Birkl2015}. To ensure proper wetting, the cells were allowed to rest for 12 hours before measurements. Cycling is conducted with a BaSyTec Cell Test System (CTS) inside an IPP750 climate chamber by Memmert operating at \qty{25}{\celsius}. For GITT, the counter electrodes are the ones from the original cell. For EIS to measure tortuosities, symmetrical cells of the negative and of the positive electrodes are constructed.  In both cases, a \qty{260}{\micro\metre} thick Whatman GF/A separator with porosity \qty{0.93} and Bruggeman coefficient \qty{1.0}{} replaces the original one, with an integrated lithium reference ring from EL-CELL for measuring the working electrode versus the reference electrode potential at \qty{0}{\volt} versus Li/Li\(^+\). For GITT, the only difference is that the original electrodes are used as counter electrodes, as shown in the SI Figure 2. The cell plungers are chosen such that the reference ring is located approximately in the middle of the separator to prevent measurement artefacts. The plungers for the EIS tortuosity measurement are copper-coated to minimize additional ohmic resistance. For GITT, \qty{120}{\micro\litre} \ce{LiPF_6} in EC:EMC:DMC 1:1:1 volume ratios (Ethylene Carbonate, Ethyl Methyl Carbonate, Dimethyl Carbonate) with 2 weight-\% VC (vinylene carbonate) from Solvionic is used to represent the original electrolyte. For EIS to measure the tortuosity of both electrodes, \qty{120}{\micro\litre} of a non-intercalating electrolyte consisting of \qty{10}{\milli\mole} Tetrabutylammonium Perchlorate (\ce{TBAClO_4}, Merck) in EC (Alfa Aesar) : EMC (Solvionic) 3:7 weight ratio is used for blocking conditions. For EIS of the separator, \qty{50}{\micro\litre} EC:EMC:DMC 1:1:1 volume ratios are used again instead. \par
The tortuosity of both electrodes and the separator is determined with EIS according to the procedure thoroughly described by Landesfeind et al. \cite{Landesfeind2016, Landesfeind2018}. EIS measurements are conducted under blocking conditions in potentiostatic mode, employing a Gamry 1010E instrument with a \qty{5}{\milli\volt} amplitude over a frequency range of \qtyrange{1}{1000}{\kilo\hertz}. To ensure measurement reproducibility, this is repeated for three cells for each component. For the impedance spectra, Equivalent Circuit Models (ECM) are used to obtain the ionic resistance \(R_\text{ion}\) from which the tortuosity is then calculated. With \(A\) denoting cross-section area and \(L_k\) denoting coating thicknesses, \(R_\text{ion}\) can be obtained according to

\begin{equation}
    \tau^2 = \frac{\varepsilon R_\text{ion} A \kappa_e}{2L_k},
\end{equation}

with the conductivity of the electrolyte at \(\kappa_e = \qty{0.32}{\milli\siemens\per\centi\metre}\) and the \(2\) referring to the fact that we have two identical coatings in the symmetrical cell. The ECM for the separator consists of a resistor \(R_{ion}^*\) in series with a constant-phase element. The ECM for the electrodes consists of a resistor \(R_{ion}\) in series with a simplified Transmission Line Model (TLM). For the latter, blocking conditions, reflective boundary conditions, and \(R_\text{ion} \gg R_\text{electrolyte}\) are assumed. See Schmitt et al. \cite{Schmitt2023} for further elaborations. An ECM consisting of a resistor and capacitor in parallel fits the impedance semicircle at \qtyrange{4}{100}{\hertz} and yields the exchange-current densities. \par
The electronic conductivities of the electrodes are determined from a four-point-probe measurement (Ossila) to be \(\sigma_n^* = \qty{215}{\siemens\per\metre}\) and \(\sigma_p^* = \qty{0.25}{\siemens\per\metre}\). An adhesive tape is used to delaminate the coating from the current collector to ensure that only the conductivity of the porous electrode is measured. \par


\subsection{Data for Results and Discussions}

\begin{figure*}[!ht]
    \centering
    \includegraphics[width=2\columnwidth]{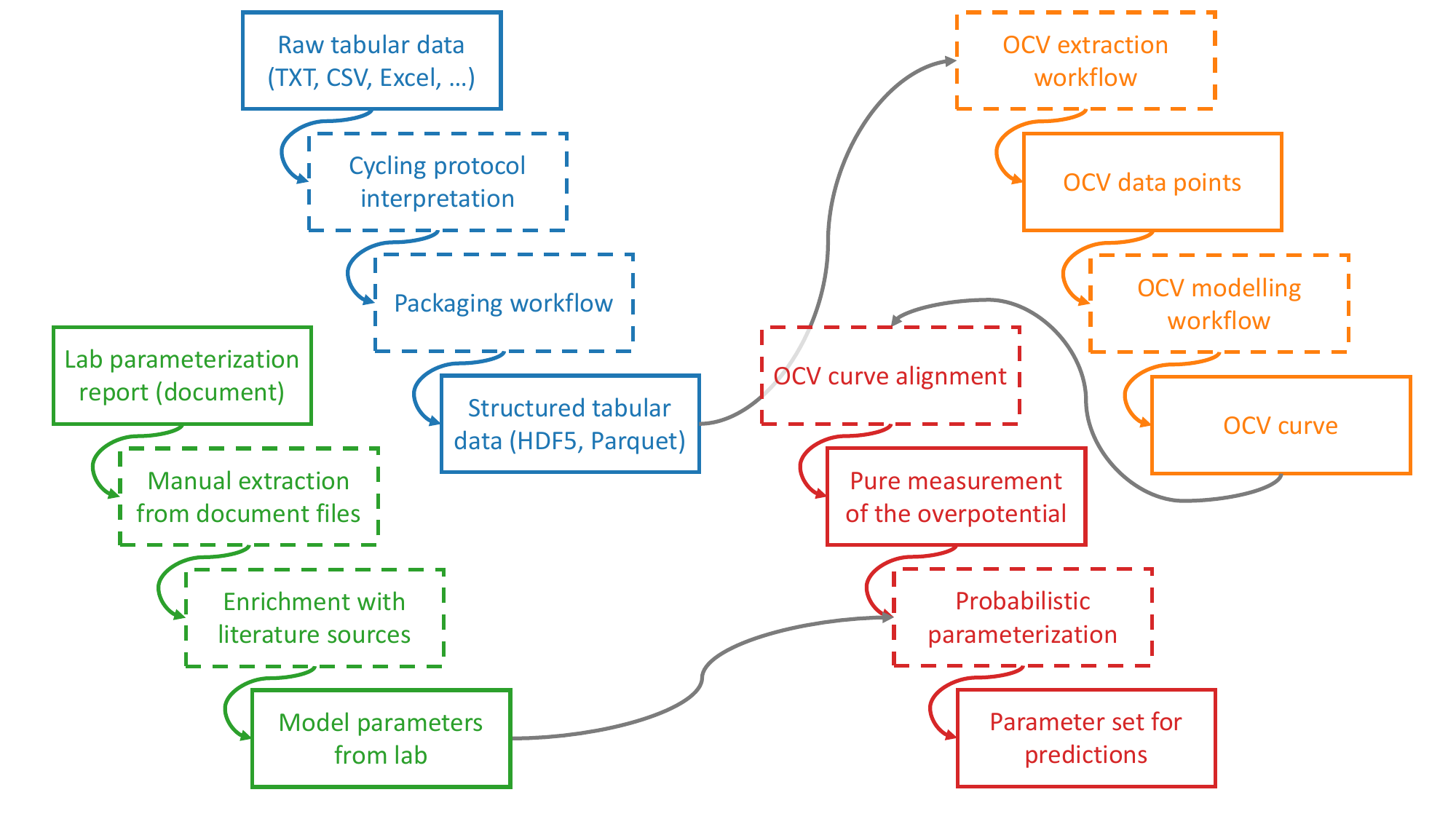}
    \caption{Finalized workflow for handling GITT and/or EIS data. Solid lines indicate a Record/dataset, while dashed lines indicate a Workflow/sub-task. The four bigger coloured sections represent the raw-to-interoperable data conversion (top left, blue), laboratory-to-interoperable report conversion (bottom left, green), discerning static and dynamic measurement features (top right, orange), and model parameterization (bottom right, red).}
    \label{fig:workflow}
\end{figure*}

We emulate the state-of-the-art of the research field by considering data from one of our previously finished experiments, which had little interaction between experimentalists and theoreticians. We use this as a basis for discussion between the two disciplines and to motivate how a more FAIR-compliant version of the experiment and its interpretation improve results. This dataset comprises GITT and EIS measurements from a Basytec device setup on the EL-CELL setup. Three full charge-discharge cycles at C/5 current between \qty{2.5}{\volt} and \qty{4.2}{\volt} were performed before each measurement. The pulse charges are given relative to a scale between 0 \% and 100 \% corresponding to \qty{2.5}{\volt} and \qty{4.2}{\volt} at C/50 current cycling, respectively. Between 10 \% and 90 \% SOC, the GITT pulses were carried out with C/10 current in 5 \% steps, and beyond that with C/20 current in 1 \% steps to avoid reaching cut-off voltages early and get a higher resolution at the edges. The relaxation criterium signalling the end of the rest phases is a voltage change smaller than \qty{0.0005}{\volt} within the last 30-minute segment.
The GITT measurements and EIS measurements were performed interspersed with each other. The precise timings and order of operations are collected in the SI Table 1 and the SI Table 2 for the lithiation and delithiation direction, respectively. \par
Our discussions revealed the necessity of a GITT measurement without performing EIS intermittently for parameterization. The exact measurement protocol is listed in the SI Table 3 and the SI Table 4 in lithiation and delithiation direction, respectively. \par
To handle the measurement data, we first need to convert it into a consistent format suited to our analysis. The battery measurement devices (\textquote{cyclers}) commonly output data in a proprietary format.
The most raw export from the cyclers is a .csv file.
We stripped redundant measurement columns, like various representations of time or empty columns, reducing file size by 77 \%.
We packaged the remaining table in a Parquet file, reducing file size by another 87 \%.
Parquet is a minimal file size container, exploiting common redundancies in time-series data.
Parquet also provides fast access with its column-oriented data structure. \par

\section{Results}

Our developed workflow is summarized in \prettyref{fig:workflow}. We group its components into four categories: converting raw measurement data into interoperable data (highlighted in blue), collating laboratory reports into interoperable characterization results (highlighted in green), distinguishing static from dynamic measurement features (highlighted in orange), and parameterizing an electrochemical model (highlighted in red). We now discuss these in order. \par

\subsection{Creating interoperable data}

We aim to standardize data processing by establishing one consistent data format internally. With this approach, reusing our existing data processing scripts for future datasets becomes seamless —- requiring only a single script each time to convert new datasets into the standardized format, or maybe even just different settings in the same script. We showcase our data conversion on the .csv conversion of our proprietary cycler output file, as .csv is the most generally accessible non-proprietary format.
The first standardization here deals with general tabular data interpretation regardless of format, e.g., conversion to SI units or structuring according to measurement protocol. \par
The development of such an interpretation script against multiple datasets revealed the following adjustments it needs to be able to perform.
\begin{itemize}
    \item Normalizing column descriptors, e.g., from \textquote{Applied current  /  \unit{\ampere\per\square\metre}} to \textquote{I [\unit{\ampere}]}; the re-formatted unit denotion is intentional, and ambiguous extra scalings like the \unit{\square\metre} obfuscate magnitudes.
    \item Stripping redundant or empty data columns, as these would slow down network-intensive data processing and obscure the information content.
    \item Stripping superfluous data rows, e.g., measurement channels that were logged even though no experiment was connected to them.
    \item Storing non-data comments in a separate file.
    \item Converting the file encoding into a global format.
    \item Converting localized column delimiters and decimal symbols into a global format.
    \item Interpreting the contents of a \textquote{cycler state} column of the user's choice, based on state changes of which the measurement will be segmented.
    \item Collating multiple measurement files into one while preserving the numbering of the original files for consistency.
    \item Normalizing current sign conventions based on cycler state, as some cyclers might imply current sign change by stating \textquote{Charge} and \textquote{Discharge}, while others will additionally explicitly denote it.
    \item Normalizing voltage sign conventions globally, as the direction of the battery in the cycler should not affect further data processing.
    \item Normalizing the sign convention for the imaginary part of an impedance measurement, as only some cyclers will report the true impedance. In contrast, others already negated the imaginary part for Nyquist plots.
    \item Lastly, normalizing current and voltage signs to align with the convention of the battery model and extract the working conditions to input into the battery simulator.
\end{itemize} \par
To store the standardized measurement, we use the file format Apache Parquet, as it features an optimally small file size, fast reads, and broad software support. The conversion to a Parquet file happens with one central script to ensure that the structure of the files is consistent. We showcase the general tool usage on this interpretation step in the SI Section 3. We document unstructured metadata, like the operating states of each segment, via a separate JSON file. Still, the reusability of our data processing scripts requires us to keep them file-agnostic, as they should work with minimal adjustments for data stored as CSV or HDF5. Therefore, we handle in-memory sharing between data processing scripts with a Python object structure. Reusing our data processing scripts with different file formats involves writing one script that parses them into that Python object structure. \par

\subsection{Challenges in creating interoperable laboratory reports}

We want to be able to handle any information on material and cell properties programmatically. Then, all further data processing steps document exactly how we used that information, and future reuse can build on that. We showcase our laboratory report standardization on the documents as they come, as this elucidates some of the common challenges in the communication and data exchange between experimentalists and theoreticians. A more ideal setup than what we show here would be to introduce ontology-based checklists and data sheets to the laboratory. \par
The laboratory report was summarized into two files that act as interfaces to users. The experimentalist side intended to curate the data in a self-explanatory way and devised the following attempt to structure and describe the information and results from the parameterization works. The first file, an Excel file, contained all geometry and material parameters in a minimal format. The second file, a PowerPoint file, repeated some of the parameters while also giving error bars, additional context on the methods used, and diagrams for the non-scalar information and images from the microscopy measurements. The PowerPoint file acts as metadata for the data in the Excel file. Nevertheless, the Excel file is not interoperable, as there is no machine-readable information about its data structure. Manual extraction is the only avenue here; ideally, the experimentalists would have been provided with an interoperable structure to input. Such a structure would also have given the theoreticians a central document to align their requirements to ensure that the experiments cover all required material properties. As an example, the thermodynamic factor of the electrolyte was initially missed on both sides. We fill such gaps with literature data gleaned from LiionDB \cite{Wang2022}. \par
As standardization efforts are still rapidly developing (such as the BPX physics-based battery modelling standard), we opted not to adopt them during our methodology's long development period. However, once all the required features are present, BPX will be the most interoperable way to present our parameter file. Alternatively, our parameter file is a Python script that stores the parameters in a key-value structure. The keys correspond to the simulation software PyBaMM \cite{Sulzer2021}. \par

\subsection{Data preprocessing to enhance signal interpretability} \label{subsec:static_vs_dynamic}

We want to dissect our data into signal and background. More accurately, we only want to use the part of the data containing a signal for a specific parameter of interest. Then, the sensitivity and precision of the following parameterization step are much more easily assessed. The time-series voltage response of a battery can be split into a static part (OCP), and a dynamic part, termed \textquote{overpotential}. Since we want to study transport properties, which only appear in the dynamic part here, we must first consider the static part. \par
We extract OCP data of both electrodes from GITT measurements as described in \ref{subsec:ElectrochemicalTheory}.
We store the extracted OCV data in a JSON file; as this dataset is rather small, JSON is more appropriate here than Parquet, for human-readability.
To increase the SOC resolution and filter noise, we interpolate the OCP data with the OCP model of Birkl et al. \cite{Birkl2015}. These steps entail many small adjustments and a carefully crafted optimization algorithm, which are documented by our code and the workflow files describing its invocation. See Yao et al. \cite{Yao2024} for another example. Finally, we store the OCV model and the metadata of its optimization in a JSON file. The data, the fit parameters, a directly usable representation of the fit function, and the optimization metadata are stored with respective keys. \par
We want to parameterize our data in a way that considers as many uncertainties as possible, as battery measurements, in particular, entail a lot of them \cite{Kuhn2022}. Then, our results will transparently encode how accurately the battery response reflects its material properties and allow us to update the range of possible parameters with future measurements. First, we subtract the optimized OCP model from the GITT data.
To verify the accuracy and alignment of the OCP model, we plot the resulting overpotential measurement and check that the voltage asymptotes of the rest phases are close to zero.
The overpotential is stored as a Parquet file with an identical internal structure to the original data, including timestamps, current, and voltage. \par

\subsection{Probabilistic parameterization}

We now prepare and perform the parameterization according to the algorithm EP-BOLFI \cite{Kuhn2022}. EP-BOLFI splits into a preprocessing step for Expectation Propagation (EP) and a parameterization step for Bayesian Optimization for Likelihood-Free Inference (BOLFI). The application to GITT is part of the EP-BOLFI publication. Preprocessing for EP allows you to apply domain knowledge by transforming the data into characteristic features. A typical GITT pulse or rest phase for materials like graphite and NMC can be entirely described by only two features, if no phase changes are occurring during the measurement, comprising of a total of five scalar values: the square-root behaviour for short times consisting of offset and square-root slope, and the exponential behaviour for long times consisting of offset, magnitude, and decay rate. \cite{Kuhn2022} We choose a suitable subset of features that relate to the quantity we wish to measure; here, it is the short-time square-root slopes for diffusivities. The remaining central input EP-BOLFI requires are our prior assumptions about the parameters of interest. The spread of sensible parameters that is known a priori is encoded as a probability distribution, denoted as the Prior. \par
All inputs for the parameterization get encapsulated as a JSON file containing model information, model discretization, experimental conditions, experimental data, experimental features, and EP-BOLFI settings. We now visualize that our parameterization is set up correctly.
We do so by collecting the spread of simulation results over the parameter sets that are at the 95 \% probability bounds of the Prior.
After visually confirming that the Prior we set contains the true parameters in its 95 \% probability bounds, we run the parameterization. See the GITT analysis in the EP-BOLFI paper \cite{Kuhn2022} for a detailed explanation of this process. \par
The parameterization result is also a probability distribution over the parameter sets. Compared to the Prior, it only contains the subset of the Prior that also agrees with the data. As the result is a posteriori knowledge about the true parameters, it is aptly denoted the Posterior.
We can visualize the Posterior similarly to the Prior, as it is structurally identical. Hence, the plots are consistent and can emphasize that the Posterior is a knowledge update of the Prior. Once the parameterization is done for each GITT pulse, we collect the individual SOC point parameters into a function of SOC.
The SOC-dependent functions are stored as JSON files alongside a B-spline interpolation in Python format and their plot. \par

\subsection{GITT characterization results}

\begin{figure}[!ht]
    (a) \vspace*{-0.6em}\begin{center}
        \includegraphics[width=\columnwidth]{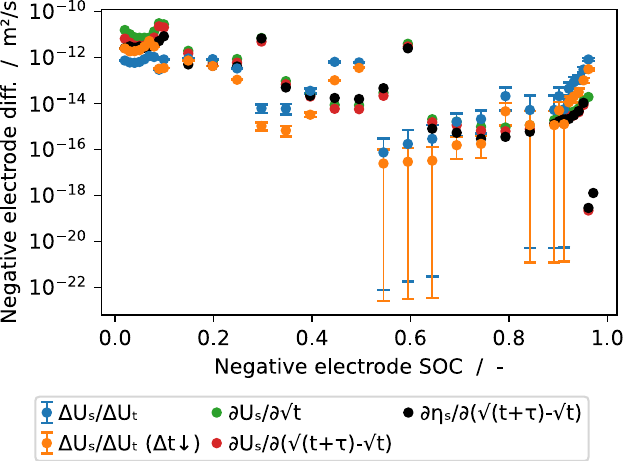} \\
    \end{center}
    (b) \vspace*{-0.6em}\begin{center}
        \includegraphics[width=\columnwidth]{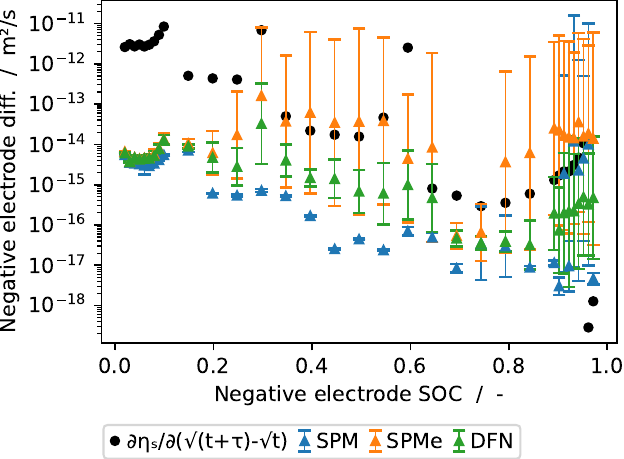}
    \end{center}
    \caption{Results for the diffusivity of the active material from one set of GITT data in delithiation direction, via direct calculations (a) and from fitting electrochemical models (b). The labels read as follows: \(\Delta U_s / \Delta U_t\) refers to the original GITT method \cite{Weppner1977}, \(\Delta U_s / \Delta U_t (\Delta t \downarrow)\) refers to the same method applied to only a suitably small time segment (\qty{90}{\second}), \(\partial U_s / \partial \sqrt{t}\) refers to the differential formulation of the original GITT method, \(\partial U_s / (\partial \sqrt{(t + \tau} - \sqrt{t})\) refers to a correction for overlapping relaxation phenomena \cite{Kang2021}, \(\partial \eta_s / (\partial \sqrt{(t + \tau} - \sqrt{t})\) additionally removes the OCP prior to diffusivity calculation, and SPM, SPMe, and DFN refer to the fitted electrochemical models. The best direct approach is plotted in black in (b) as well for comparison.}
    \label{fig:gitt-diffusivity-delithiation}
\end{figure}

\newpage

\prettyref{fig:gitt-diffusivity-delithiation}a shows the results of state-of-the-art direct diffusivity extraction from the GITT data in delithiation direction. The limited error propagation we can consider here only displays the effect of voltage measurement resolution. It naturally becomes an issue in the SOC range \qtyrange{0.6}{1.0}{}, where graphite has a voltage plateau that is shallower than the measurement can resolve. Hence, we observe large errorbars in that range. \par

\prettyref{fig:gitt-diffusivity-delithiation}b shows the results of our model-based diffusivity extration from the same GITT data in delithiation direction. Our approach includes more sources of uncertainty in its error propagation, especially parameter uncertainties and their correlations. We observe a significant decrease in diffusivity accuracy at a much wider SOC range \qtyrange{0.3}{1.0}{}. \par

\begin{figure}[!ht]
    \centering
    \includegraphics[width=\columnwidth]{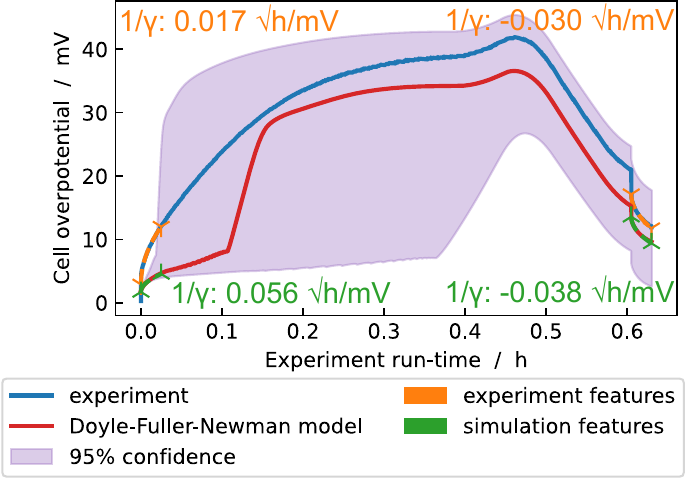}
    \caption{The predictive parameterization posterior of a GITT measurement in delithiation direction. The highlighted square-root slopes \(\gamma\) are used for fitting. The constant-current pulse lasts \qty{0.6}{\hour}, and we show only the relevant part of the following rest. The square-root features used for parameterization are noted down for experiment (orange) and optimal simulation (green) in \unit[power-half-as-sqrt]{\second\tothe{0.5}\per\volt}. The large posterior 95 \% confidence interval is a consequence of the non-matchable pulse square-root feature.}
    \label{fig:gitt_parameterization_delithiation}
\end{figure}

\begin{figure}[!ht]
    (a) \vspace*{-0.8em}\begin{center}
        \includegraphics[width=\columnwidth]{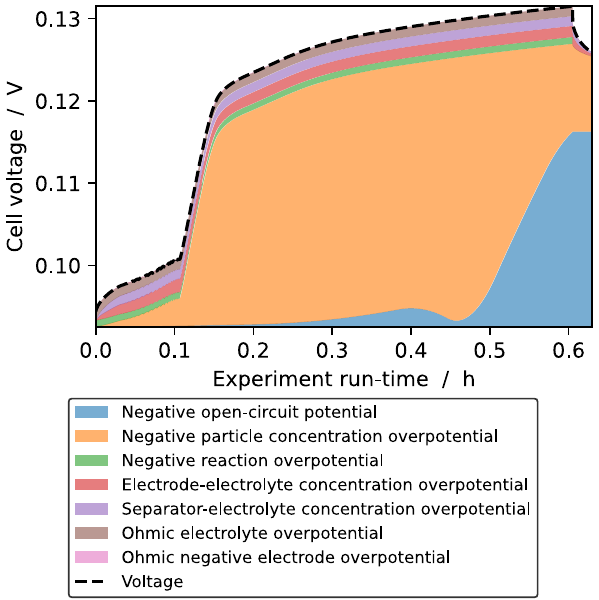}
    \end{center}
    (b) \vspace*{-0.8em}\begin{center}
    \includegraphics[width=\columnwidth]{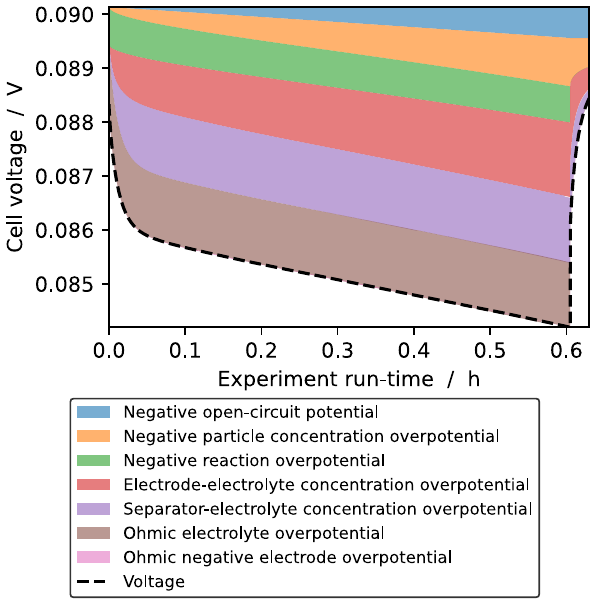}
    \end{center}
    \caption{A plot detailing the overpotential components in a GITT measurement in delithiation (a) and lithiation (b) direction. Only the two largest contributions are relevant in the delithiation direction, which are the OCP and particle concentration overpotential. The oscillation between the two is a result of a rapid change in OCP slope. All contributions are equally important in the lithiation direction. In particular, we see that the particle concentration overpotential shows a minor contribution overall, which makes this a measurement of the electrolyte rather than of the electrode.}
    \label{fig:gitt_parameterization_components}
\end{figure}

The DFN simulations for the individual GITT pulses, where one exemplary one is shown in \prettyref{fig:gitt_parameterization_delithiation}, hint at the reason. Traditional GITT relies on the assumption that the overpotential response grows monotonously, which we do not observe there. We investigate the unexpected shape of the overpotential further in an analysis of the overpotential components in \prettyref{fig:gitt_parameterization_components}a. \par

Oscillations between SOC and overpotential occur, showing unexpected retrograde SOC change. The SOC at which this happens is near a kink in the OCP, originating from crystal structure rearrangements in the active material. While the voltage response grows monotonously, the overpotential response does not, which would be missed in traditional GITT. With our approach, though, the similarity of the shape and the relaxation square-root accuracy tell us that the OCV and model accuracy are sufficiently high for parameterization. \par

Furthermore, we observe a fundamental phenomenon in statistical estimation in the model-based diffusivities \prettyref{fig:gitt-diffusivity-delithiation}b: the bias-variance tradeoff. Since the SPM neglects electrolyte effects, it is the wrong model and can not fit the data, which we call a high bias. Consequently, as seen from the error bars, the variance is suspiciously low, which we colloquially call \textquote{confidently incorrect}. As we approach a more correct model with the SPMe, the variance grows, which we now know is expected but may be counterintuitive. \cite{Escalante2021} Only the DFN, as a sufficient model, can exhibit low bias and variance simultaneously. This example cautions us to trust a parameterization with a single model without considering the context from adjacent models. \par

\begin{figure}[!ht]
    (a) \vspace*{-0.6em}\begin{center}
        \includegraphics[width=\columnwidth]{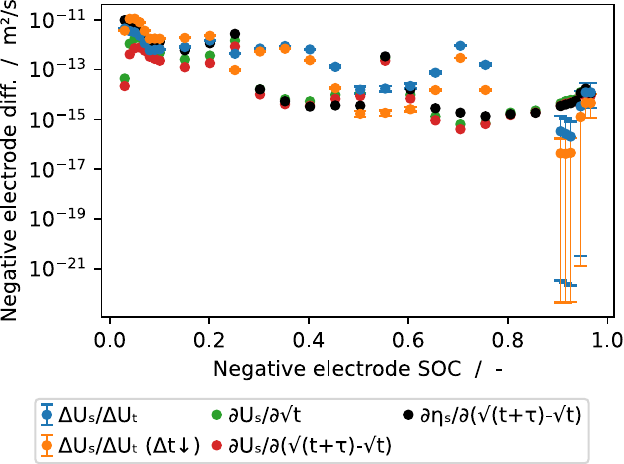} \\
    \end{center}
    (b) \vspace*{-0.6em}\begin{center}
        \includegraphics[width=\columnwidth]{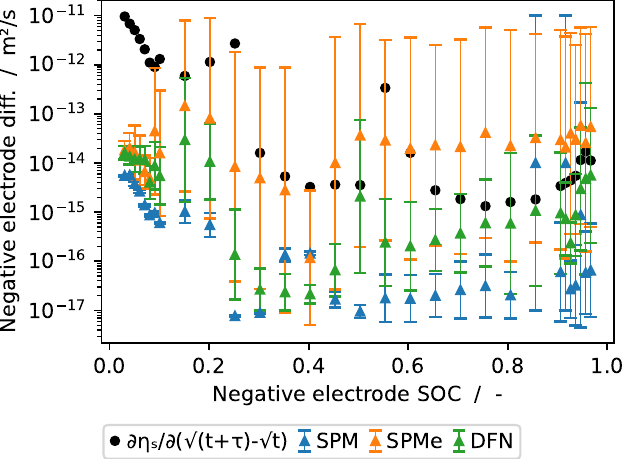}
    \end{center}
    \caption{Results for the diffusivity of the active material from one set of GITT data in lithiation direction, via direct calculations (a) and from fitting electrochemical models (b). The labels read as follows: \(\Delta U_s / \Delta U_t\) refers to the original GITT method \cite{Weppner1977}, \(\Delta U_s / \Delta U_t (\Delta t \downarrow)\) refers to the same method applied to only a suitably small time segment (\qty{90}{\second}), \(\partial U_s / \partial \sqrt{t}\) refers to the differential formulation of the original GITT method, \(\partial U_s / (\partial \sqrt{(t + \tau} - \sqrt{t})\) refers to a correction for overlapping relaxation phenomena \cite{Kang2021}, and \(\partial \eta_s / (\partial \sqrt{(t + \tau} - \sqrt{t})\) additionally removes the OCP prior to diffusivity calculation, and SPM, SPMe, and DFN refer to the fitted electrochemical models. The best direct approach is plotted in black in (b) as well for comparison.}
    \label{fig:gitt-diffusivity-lithiation}
\end{figure}

\prettyref{fig:gitt-diffusivity-lithiation}a shows the results of state-of-the-art direct diffusivity extraction from the GITT data, this time in the lithiation direction. The limited error propagation we can consider here again only displays the effect of voltage measurement resolution. This time, it is much less of an issue due to the measurement happening in the direction of increasing OCP slope, which results in voltage responses comfortably beyond measurement accuracy. The exception is at the very beginning of the GITT lithiation measurement at high negative electrode SOC, as the surface concentrations do not reach the non-plateau region of the OCP yet. \par

\prettyref{fig:gitt-diffusivity-lithiation}b shows the results of our model-based diffusivity extraction on the same GITT data in lithiation direction. We observe almost no improvement over the prior parameter assumptions for the SOC range \qtyrange{0.3}{1.0}{}. The SPM fit shows suspiciously low variance, which can be attributed to an insufficient model introducing significant bias \cite{Escalante2021}. We see a marked decrease in diffusivity accuracy in the SOC range \qtyrange{0.0}{0.1}{} across all models this time. Similar to the delithiation direction, we observe that GITT measurements towards the edge of the SOC range can not uniquely parameterize the active material diffusivity. When the local electrode concentration hits an SOC limit, a \textquote{depletion shockwave} runs from the current collector to the separator, which has a different dynamic than a diffusivity response. \par

\begin{figure}[t]
    \centering
    \includegraphics[width=\columnwidth]{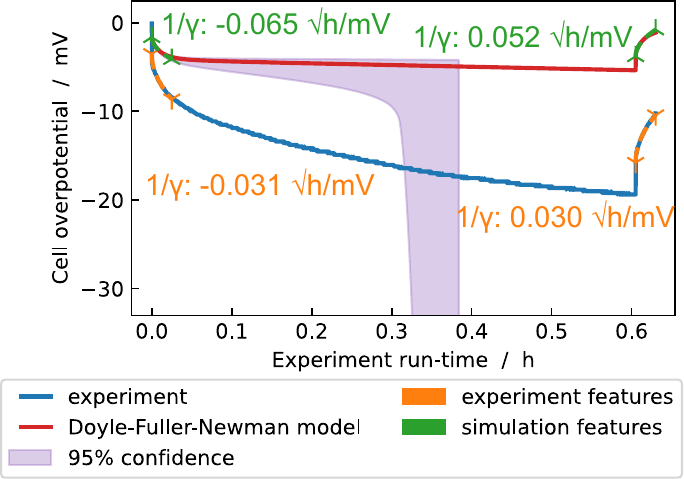}
    \caption{The predictive parameterization posterior of a GITT measurement in lithiation direction. The highlighted square-root slopes \(\gamma\) are used for fitting. The constant-current pulse lasts \qty{0.6}{\hour}, and we show only the relevant part of the following rest. The square-root features used for parameterization are noted down for experiment (orange) and optimal simulation (green) in \unit[power-half-as-sqrt]{\second\tothe{0.5}\per\volt}. The overfitted posterior 95 \% confidence interval is a consequence of the prior 95 \% confidence interval not enveloping the data either.}
    \label{fig:gitt_parameterization_lithiation}
\end{figure}

The DFN simulations can't capture the magnitude of the overpotential this time, as we see in one of the parameterized pulses in \prettyref{fig:gitt_parameterization_lithiation}. \par

The shallow OCP curve is one reason, as the negative electrode concentration overpotential scales with it. Consequently, it is small with regards to the electrolyte overpotential, as shown in the overpotential analysis in \prettyref{fig:gitt_parameterization_components}b. \par

We repeat the identical procedure for the positive NMC electrode in the SI Section 4. As NMC has a benign OCP with no kinks and small slope changes, traditional GITT works well and our approach is not needed. \par

We can ensure the compatibility of other measurements to our GITT parameterization by utilizing the fact that we treated it according to Bayesian principles. In Bayesian statistics, results from insufficient data are described as probability distributions reflecting the uncertainty that the data and model contain.
To give context: the posterior (the \textquote{result}) in Bayesian statistics is obtained as the product of the  prior (the \textquote{informed researcher's intuition}) and the likelihood (the \textquote{model}). We generalize this multiplication update by multiplying the likelihoods of another measurement and GITT by the prior. The mathematical justification for this \textquote{trick} stems from summary statistics \cite{Minka2001}. More straightforwardly, this is equivalent to a semi-parallelized EP-BOLFI, which is stated by Barthelmé et al. \cite{Barthelme2015} to be valid. This \textquote{simple} step is only possible when the models used for the parameterization in both cases are compatible. See Zhu et al. \cite{Zhu2017} and Deng et al. \cite{Deng2020} for the challenges that combining GITT and, e.g., Electrochemical Impedance Spectroscopy otherwise entail.

\section{Discussions}

Here we describe the generally applicable issues and improvements we found in our collaboration. We will discuss these in order: measurement protocol communication, measurement objective communication, measurement accuracy assessment, uncertainty treatment, documentation via metadata, elucidating domain knowledge, checking model compatibility, interoperable laboratory reports, and software dependency review. \par
Communication of measurement protocol is the first step that may induce issues. We found different understandings of a measurement technique (GITT) amongst the parties involved. With different requirements and limitations, one may select a different interpretation. For example, a theoretician might prefer a rigid GITT set for consistency or an arbitrary, but uninterrupted one for error mitigation. At the same time, an experimentalist is concerned about maximizing expensive equipment time and might interweave other measurements in the rest phase \textquote{downtime}. By joining all parties on what may be considered the domain of only one party, we could find an optimal solution for all: shortening the rest of the phases. While shorter rest phases down to \qty{15}{\minute} may suffice\cite{Kuhn2022}, this only applies after one verification GITT pulse on the material at hand with the usual hours-long relaxation. \par
Measurement objective communication is a separate step that needs to be considered. The issue we found was different quantities of interest. This may sound trivial without knowledge of experimental setups, but to use resources optimally, they can be much more complex than what their output files suggest. For example, a multiplexer setup can seamlessly switch between the time-domain measurements for GITT and the frequency-domain measurements for EIS. However, specific quantities can only be logged by one device at once, leading to gaps in the record for the other device. For example, an EIS measurement device may not be set up to track the total charge transferred, which is necessary for the GITT measurement device to assign SOCs to data points. A solution to avoid this is to agree on a verbose spreadsheet with exact cycler instructions beforehand. Some issues only appear in such simplified discussions, as they eliminate the application of advanced knowledge. For example, a theoretician might not know the time, current, or voltage resolution limits. \par
Measurement accuracy assessment refers to a human-interpretable representation of the intermediate steps in the data pipeline. On the one hand, it allows for the re-calibration and fine-tuning of the intermediate steps. On the other hand, it reduces the individual errors that accumulate in the final error propagation calculation. The issue we found was a lack of checks of assumptions. For example, a GITT measurement may be idealized as per theory. Each segment starts with a short-term square-root behaviour and smoothly merges into an exponential decay towards a (quasi-)equilibrium. To verify this, we subtracted the electrode OCP from the data. But we found oscillations of the overpotential around kinks in the electrode OCP, e.g., in \prettyref{fig:gitt_parameterization_delithiation}. While it is well understood that the original GITT formula from 1977 \cite{Weppner1977} does not apply in such situations, we show that GITT with a model-based analysis can still yield a suitable parameterization. \par
Uncertainty treatment is a step that can not be overstated in its importance in battery research. We found that the magnitude of uncertainty sources is easily underestimated when only considering one at a time. With EP-BOLFI \cite{Kuhn2022}, we turn to a black-box optimizer with a stochastical framework that allows us to evaluate any uncertainties simultaneously that we can incorporate into a simulation model. Voltage measurement precision and material/geometrical property uncertainty can be tacked onto any simulation model. Meanwhile, material/geometrical property correlation is an intrinsic property of the model equations that EP-BOLFI uncovers. For example, the influence of electrolyte properties and their geometry on the complete parameterization is often underestimated. To verify the extent of this influence, we perform the overpotential analyses in \prettyref{fig:gitt_parameterization_components} for selected SOC points in both the delithiation and lithiation directions. We see that in delithiation direction, most of the signal stems from the electrode concentration gradients, which is desired. But in lithiation direction, only about 10-20 \% of the signal stems from the phenomenon of interest, while the electrolyte concentration gradient effects dominate the signal. Any uncertainty in the electrolyte properties has a proportionally increased influence on the parameterization of the active material diffusivity. Checking the influence of the electrolyte this way tells us how much we need to optimize the experimental setup for a sufficient signal from the electrodes. \par
Documentation via metadata is often considered an ungrateful task, as it is thought not to have an immediate benefit. The issue we found is that domain-specific language between experimentalists and theoreticians did not diverge in the words used but in the meaning of those words. For example, the term \textquote{tortuosity \(\tau\)} has different defaults depending on one's own research field. More accurately, one may refer to \textquote{path-length tortuosity \(\tau\)} if the ratio between material and effective transport properties is \(\varepsilon / \tau^2\), or \textquote{effective tortuosity \(\tau\)} if it is \(\varepsilon / \tau\). With ontology-backed descriptions of each use of the term \textquote{tortuosity}, one could directly translate sources from other disciplines into one's own requirements. \par
Elucidating domain knowledge is critical for successful communication across disciplines. The issue we found is the unconscious application of domain knowledge. For example, initial communication about the difference between a commercial cell and its modified experimental sample was kept \textquote{simple} in the interest of each party`s time: the theoreticians got the description that the experimental setup minimizes \textquote{the effect} of the separator. But this \textquote{simple} statement encodes a large volume of expectations of the measurement, an assumption on the quantities that will be extracted from it, and the method by which the separator was made \textquote{negligible}. Theoreticians would assume that the new separator would be a marginally thick glass fibre with unity tortuosity. We show the actual picture in the SI Figure 2. The separator is \textquote{removed} from the measurement by it having unity tortuosity and high porosity. But, as the commercial electrolyte influences the signal greatly \cite{Kuhn2022} compared to a purely academic cell, combined with the considerable thickness of the separator, the removal is imperfect, which must be communicated back and forth. Hence, we recommend graphical communication as a way to transfer domain knowledge. \par
Checking model compatibility thoroughly by checking assumptions between models can be arbitrarily difficult. The issue we found was the pragmatic reliance on the fact that different models of one phenomenon are supposed to approximate the same physical reality. For example, while Transmission Line Models claim that they reproduce a Finite Volume discretization of porous microstructures, from the differences found in 1D+1D impedance simulations \cite{Hallemans2024}, we can infer that TLM model parameters do not map onto those of 1D+1D models. \par
Interoperable laboratory reports may seem like an extra step on top of the measurement documentation. The issue we found is the loss of auxiliary information, e.g., meanings of data column descriptors, differences between a battery cell and its sample for measurement, known noise sources, or even specifics of the preceding equipment use. With ontologies, we have a tool to make checklists and input masks for metadata, automatically converting laboratory notes into a complete picture. For example, we encountered data segments from a multiplexer setup that were not interpretable independently. One device seemed to have arbitrary gaps in voltage data. But both devices were active simultaneously, passing their electrical connection to the battery cell back and forth. Another example is the common description of current via \unit{\ampere\per\square\metre}, which, out of context, lacks the information if the area it refers to is the total surface area of one of the electrodes or the cross-section of one of the electrodes, and which electrode it refers to. One issue we want to emphasize is human error when reading out non-machine-readable laboratory reports. Data loss may be just a matter of not scrolling down an Excel sheet or missing that it has tabs. \par
Software dependency review entails not only the log of software versions used but, more importantly, the effects each piece of software has on the workflow. The issue we found is the sometimes non-interoperable implementation of file formats. For example, Pandas is a popular Python library for handling tabular data, offering data export into the HDF5 file format. While it is possible to store interoperable data this way \cite{Paulson2021}, the default behaviour is a file that can only be reasonably read by Pandas. This raises an unnecessary barrier to future reuse. In the worst case, one must find the same Pandas version and make it run on their system. Therefore, we recommend verifying the standard adherence of your data by opening it in a \textquote{third} software, as in, one neither party initially used. Additionally, the choice of file format depends on the size of one's organization. To show that HDF5 is appropriate for organizations with dedicated resources for data curation, we refer you to Moradpour et al. \cite{Moradpour2024}. With no dedicated resources, a file format is preferable that can not be misconstrued as the highly flexible HDF5 can. For example, we choose Apache Parquet because it forces us to organize our data in a single table each time. For unstructured data, we choose JSON, since it sacrifices file compactness for structural simplicity and universal readability. \par

\section{Conclusions}

We conclude that improving adherence to the FAIR principles and refactoring data treatment to be automatable improve transparency and reusability of the data and the software that interprets it while simultaneously allowing us to bridge disciplines. We summarize our findings by FAIR principle. \par
Findable is the principle with which we identify our first issue, the low availability of appropriate data.
The emphasis lies on \textquote{appropriate}, as in, the data you find has to match the material you are studying.
For example, there are numerous incongruent material characterizations for \textquote{graphite} \cite{Ecker2015, Schmalstieg2018, Tang2019, Doyle1996, Levi1997, Levi2003, Prada2012, Smith2006, Kim2011, Doyle2003, Srinivasan2004, Liebig2019, Cabanero2018, Persson2010, Verbrugge1999, Guo2007, Li2012, Arora1999, Perkins2012, Jiang2016, Mastali2016, Kumaresan2008, Chaouachi2021}; this exhaustive list was generated with LiionDB \cite{Wang2022}. Here, the main cause is that several different configurations of graphite are commonly referred to as just \textquote{graphite}, independent of structure, morphology, or crystallinity. The other cause is the wide range of values one gets from different parameterization approaches, not all of which may take into account the effects from voltage plateaus and their transitions. \par
Accessible is the principle with which we identify our second issue, the low accessibility of raw data.
The usual way to \textquote{use} data is to extract it from a plot of the data manually.
This greatly obscures phenomena on vastly differing timescales, as in virtually all battery phenomena.
Storing the measurement files instead of just their plots on a data publisher like Zenodo provides raw data. \par
Interoperable is the principle with which we identify our third issue, proprietary data formats.
Each manufacturer of battery cyclers has their own format.
With incomplete documentation of the settings used in a particular measurement, retrieving the data completely can become challenging.
Providing a data format and the code that can handle it allows retrieval of the original measurement data. \par
Reproducible is the principle with which we identify our fourth issue, the failure to apply novel methods.
Without a completely documented example it can be impossible to identify whether the method does not apply or if one implemented it wrong.
Ultimately, the novel method may get dropped in favour of an obsolete but established one. We observe this in the still widespread use of the \(\Delta U_s / \Delta U_t\) GITT variant over the \(\partial \eta_s / \partial(\sqrt(t + \tau) - \sqrt{t})\) one or our inverse modelling approach.
We document our analysis by providing the raw data and the code analyzing it Open Source. \par
Reusable is the principle with which we identify our last issue, one of a purely legal nature.
It is common to forget that measurement data is, by default, protected by copyright, depending on your jurisdiction.
So, if no contract clauses or other legal documents like licenses have been prepared, when the researcher who curated the data has left the field, that data is now unusable.
To protect themselves against legal proceedings, institutes and industry will not use that data, even if that is not what the researcher intended.
A consistent workflow protects against such bureaucratic issues. \par
Automation has benefits separate from the FAIR principles. Segmentation into sub-tasks allows for flexible tool adaptation and reuse for new tools.
Automation is thus accessible to an ontology treatment, enabling translations between disciplines.
The automation forces any manual adjustments to be documented, leading to a machine-readable version of any intermediate step for future use.
Undocumented adjustments vanish when one commits to generally usable sub-tasks.
The workflows are also more human-readable.
Every intermediate step is inspectable for error correction.
The whole data pipeline or parts of it can be quickly re-run with different settings.
Incidentally, the data pipeline's clear structure makes it much more user-friendly. This clarity may also bear benefits for individual datasets.
Our method is transferable and adaptable to all collaborative battery material science. \par

\section*{ORCID}

\urlstyle{same}
\noindent
Yannick Kuhn \includegraphics[height=\baselineskip]{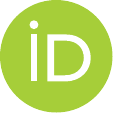} \url{https://orcid.org/0000-0002-9019-2290} \\
Bhawna Rana \includegraphics[height=\baselineskip]{Orcid_logo.pdf} \url{https://orcid.org/0000-0001-6403-5139} \\
Micha Philipp \includegraphics[height=\baselineskip]{Orcid_logo.pdf} \url{https://orcid.org/0009-0002-8705-2059} \\
Christina Schmitt \includegraphics[height=\baselineskip]{Orcid_logo.pdf} \url{https://orcid.org/0000-0002-4095-7611} \\
Roberto Scipioni \includegraphics[height=\baselineskip]{Orcid_logo.pdf}  \url{https://orcid.org/0000-0003-1926-421X} \\
Eibar Flores \includegraphics[height=\baselineskip]{Orcid_logo.pdf} \url{https://orcid.org/0000-0003-2954-1233} \\
Dennis Kopljar \includegraphics[height=\baselineskip]{Orcid_logo.pdf} \url{https://orcid.org/0000-0002-2228-2016} \\
Simon Clark \includegraphics[height=\baselineskip]{Orcid_logo.pdf} \url{https://orcid.org/0000-0002-8758-6109} \\
Arnulf Latz \includegraphics[height=\baselineskip]{Orcid_logo.pdf} \url{https://orcid.org/0000-0003-1449-8172} \\
Birger Horstmann \includegraphics[height=\baselineskip]{Orcid_logo.pdf} \url{https://orcid.org/0000-0002-1500-0578} \\

\section*{Author contributions}

\textbf{Yannick Kuhn}: Conceptualization, Data curation, Formal analysis, Investigation, Methodology, Project administration, Resources, Software, Validation, Visualization, Writing -- original draft, Writing -- review \& editing.
\textbf{Bhawna Rana}: Data curation, Formal analysis, Investigation, Validation.
\textbf{Micha Philipp}: Writing -- review \& editing.
\textbf{Christina Schmitt}: Data curation, Formal analysis, Investigation, Validation.
\textbf{Roberto Scipioni}: Formal analysis, Investigation, Visualization.
\textbf{Eibar Flores}: Conceptualization, Methodology, Writing -- review \& editing.
\textbf{Dennis Kopljar}: Conceptualization, Data curation, Funding acquisition, Investigation, Methodology, Project administration, Validation, Writing -- original draft, Writing -- review \& editing.
\textbf{Simon Clark}: Conceptualization, Data curation, Funding acquisition, Methodology, Project administration, Resources, Software, Visualization, Writing -- original draft, Writing -- review \& editing.
\textbf{Arnulf Latz}: Supervision, Writing -- review \& editing.
\textbf{Birger Horstmann}: Conceptualization, Funding acquisition, Project administration, Supervision, Writing -- review \& editing.

\section*{Conflicts of interest}

There are no conflicts to declare. \par

\section*{Data availability}

The dataset comprising of the raw experimental data, its preprocessed derivatives, and the whole parameterization workflow and results is publicly available on Zenodo: \url{https://doi.org/10.5281/zenodo.15407849}. The program needed to translate the parameterization workflow files into series of command line statements is part of the Open Source Kadi4Mat software: \url{https://gitlab.com/iam-cms/workflows}. The code these command line statements call which is performing the preprocessing and analysis, as well as the code neede to reproduce the figures, is collected within the Open Source Python library EP-BOLFI: \url{https://doi.org/10.5281/zenodo.6967238} or \textquote{pip install ep-bolfi}.

\section*{Acknowledgements}

This work was supported by the German Aerospace Center (DLR). The authors acknowledge support by the Helmholtz Association through grant no KW-BASF-6 (Initiative and Networking Fund as part of the funding measure \textquote{ZeDaBase-Batteriezelldatenbank}) and by the European Union's Horizon 2020 innovation program under grant agreement numbers 875527 (HYDRA) and  101103997 (DigiBatt). This work contributes to the research performed at CELEST (Center for Electrochemical Energy Storage Ulm-Karlsruhe). \par

\balance

\renewcommand\refname{References}

\bibliography{library}
\bibliographystyle{rsc}

\end{document}


\title{Supporting Information\texorpdfstring{\\}{ }Bridging Disciplines in Scientific Collaboration via Structured Automation in Battery Science}
\author{
\begin{minipage}{\columnwidth}
    Yannick Kuhn, Bhawna Rana, Micha Philipp, Christina Schmitt, Roberto Scipioni, Eibar J. F. Cedeño, Dennis Kopljar, Simon Clark, Arnulf Latz, Birger Horstmann
\end{minipage}
}
\maketitle

\tableofcontents

\newpage

\section{Experimental protocols}

\begin{table*}[!ht]
    \caption{First dataset measurement protocol in lithiation direction. \textquote{GITT} refers to the charge or discharge pulse of the GITT method only.}
    \label{tab:first_dataset_charge}
    \begin{tabular}{c|c|c|c}
    Operation & Current [mA] & Duration [h] & Repeats \\
    \hline 
    Rest & 0 & 2 & \multirow{6}{*}{1} \\
    EIS & -- & 0.58 &  \\
    Rest & 0 & 1 &  \\
    GITT & 0.5 & 0.24 &  \\
    Rest & 0 & 3.5 &  \\
    EIS & -- & 0.58 &  \\
    \hline 
    Rest & 0 & 1 & \multirow{4}{*}{5} \\
    GITT & 0.5 & 0.24 &  \\
    Rest & 0 & 5.1 &  \\
    EIS & -- & 1 / 0.58 / 0.58 / 0.58 / 0.77 &  \\
    \hline 
    Rest & 0 & 1 & \multirow{4}{*}{4} \\
    GITT & 0.5 & 0.24 &  \\
    Rest & 0 & 2 / 1.5 / 1 / 1 &  \\
    EIS & -- & 0.59 &  \\
    \hline 
    Rest & 0 & 1 & \multirow{4}{*}{9} \\
    GITT & 1 & 0.59 &  \\
    Rest & 0 & 3.5 / 2.5 / 2 / 1.5 / 1.5 / 1 / 1 / 2 / 1.5 &  \\
    EIS & -- & 0.6 &  \\
    \hline 
    Rest & 0 & 1 & \multirow{4}{*}{4} \\
    GITT & 1 & 0.59 &  \\
    Rest & 0 &  1 &  \\
    EIS & -- & 0.61 &  \\
    \hline 
    Rest & 0 & 1 & \multirow{3}{*}{3} \\
    GITT & 1 & 0.59 &  \\
    EIS & -- & 0.58 / 0.69 / 0.58 &  \\
    \hline 
    Rest & 0 & 1 & \multirow{4}{*}{9} \\
    GITT & 0.5 & 0.24 &  \\
    Rest & 0 & 1 / 1 / 1.5 / 2.5 / 5 / 6.1 / 6.1 / 5 / 6.1 &  \\
    EIS & -- & 0.58 / 0.58 / 0.61 / 0.61 / 0.61 / 0.95 / 14.85 / 1 / 1 &  \\
    \hline 
    Rest & 0 & 1 & \multirow{3}{*}{1} \\
    GITT & 0.5 & 0.24 &  \\
    Rest & 0 & 7.1 &  \\
    \end{tabular}
\end{table*}

\begin{table*}[!ht]
    \caption{First dataset measurement protocol in the de-lithiation direction. \textquote{GITT} refers to the charge or discharge pulse of the GITT method only.}
    \label{tab:first_dataset_discharge}
    \begin{tabular}{c|c|c|c}
    Operation & Current [mA] & Duration [h] & Repeats \\
    \hline 
    Rest & 0 & 2 & \multirow{6}{*}{1} \\
    EIS & -- & 0.61 &  \\
    Rest & 0 & 1 &  \\
    GITT & 0.5 & 0.24 &  \\
    Rest & 0 & 1 &  \\
    EIS & -- & 0.58 &  \\
    \hline 
    Rest & 0 & 1 & \multirow{4}{*}{6} \\
    GITT & 0.5 & 0.24 &  \\
    Rest & 0 & 1 &  \\
    EIS & -- & 0.58 &  \\
    \hline 
    Rest & 0 & 1 & \multirow{4}{*}{3} \\
    GITT & 0.5 & 0.24 &  \\
    Rest & 0 & 1 &  \\
    EIS & -- & 0.61 &  \\
    \hline 
    Rest & 0 & 1 & \multirow{4}{*}{9} \\
    GITT & 1 & 0.59 &  \\
    Rest & 0 & 1 / 1 / 1 / 1 / 1.5 / 3.5 / 1 / 1 / 1 &  \\
    EIS & -- & 0.61 / 0.60 / 1.1 / 0.60 / 0.60 / 0.59 / 0.60 / 1.1 / 0.61 &  \\
    \hline 
    Rest & 0 & 1 & \multirow{4}{*}{7} \\
    GITT & 1 & 0.59 &  \\
    Rest & 0 & 1 / 1.5 / 1 / 1.5 / 1.5 / 2 / 3.5 &  \\
    EIS & -- & 0.61 / 0.6 / 1.1 / 0.60 / 0.60 / 0.59 / 0.59 &  \\
    \hline 
    Rest & 0 & 1 & \multirow{4}{*}{9} \\
    GITT & 0.5 & 0.24 &  \\
    Rest & 0 & 4.6 / 4.6 / 4 / 5.1 / 6.1 / 6.1 / 5.1 / 5.6 / 6.1 &  \\
    EIS & -- & 0.91 / 0.59 / 0.59 / 0.60 / 0.59 / 0.59 / 1.5 / 0.58 / 0.58 &  \\
    \hline 
    Rest & 0 & 1 & \multirow{3}{*}{1} \\
    GITT & 0.5 & 0.24 &  \\
    Rest & 0 & 7.1 &  \\
    \end{tabular}
\end{table*}

\vspace*{10em}

\newpage

\phantom{placeholder text} \\

\newpage

\begin{table*}[!ht]
    \caption{Second dataset measurement protocol in lithiation direction. \textquote{GITT} refers to the charge or discharge pulse of the GITT method only.}
    \label{tab:second_dataset_charge}
    \begin{tabular}{c|c|c|c}
    Operation & Current [mA] & Duration [h] & Repeats \\
    \hline 
    Rest & 0 & 3 & 1 \\
    \hline 
    GITT & 0.5 & 0.24 & \multirow{2}{*}{10} \\
    Rest & 0 & 6.1 &  \\
    \hline 
    GITT & 1 & 0.61 & \multirow{2}{*}{16} \\
    Rest & 0 & 4 / 4 / 3.5 / 3.5 / 2.5 / 2.5 / 2 / 2 / 2.5 / 2.5 / 2 / 2 / 2.5 / 2.5 / 2.5 / 2.5 &  \\
    \hline 
    GITT & 0.5 & 0.24 & \multirow{2}{*}{10} \\
    Rest & 0 & 2.5 / 2.5 / 2.5 / 2 / 2.5 / 2 / 2 / 2 / 2 / 2 &  \\
    \end{tabular}
\end{table*}

\begin{table*}[!ht]
    \caption{Second dataset measurement protocol in the de-lithiation direction. \textquote{GITT} refers to the charge or discharge pulse of the GITT method only.}
    \label{tab:second_dataset_discharge}
    \begin{tabular}{c|c|c|c}
    Operation & Current [mA] & Duration [h] & Repeats \\
    \hline 
    Rest & 0 & 3 & 1 \\
    \hline 
    GITT & 0.5 & 0.24 & \multirow{2}{*}{10} \\
    Rest & 0 & 2 &  \\
    \hline 
    GITT & 1 & 0.61 & \multirow{2}{*}{16} \\
    Rest & 0 & 2 / 2 / 2 / 2 / 3 / 3 / 2 / 2 / 2 / 2.5 / 2.5 / 2 / 3 / 2.5 / 4.5 / 5 &  \\
    \hline 
    GITT & 0.5 & 0.24 & \multirow{2}{*}{10} \\
    Rest & 0 & 6.6 / 6.1 / 7.1 / 7.1 / 7.1 / 7.1 / 7.1 / 7.1 / 7.1 / 7.1 &  \\
    \end{tabular}
\end{table*}

\section{Experimental elaborations}

\subsection{Cell disassembly}

\begin{figure}[!ht]\begin{center}
    \includegraphics[width=0.66\textwidth]{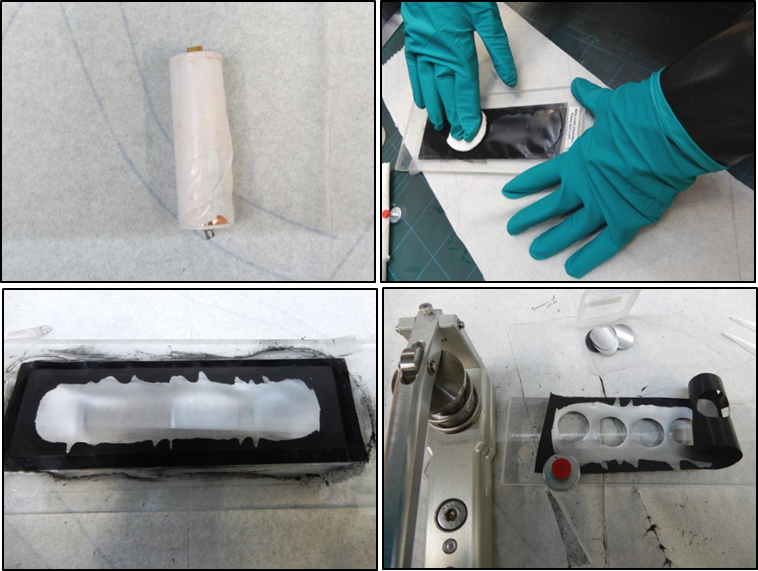}
    \caption{Photos from the experimental procedure to prepare electrodes for electrochemical analysis.}
    \label{fig:cell_disassembly}
\end{center}\end{figure}

\subsection{Cell reassembly}

\begin{figure}[!ht]\begin{center}
    \includegraphics[width=0.75\textwidth]{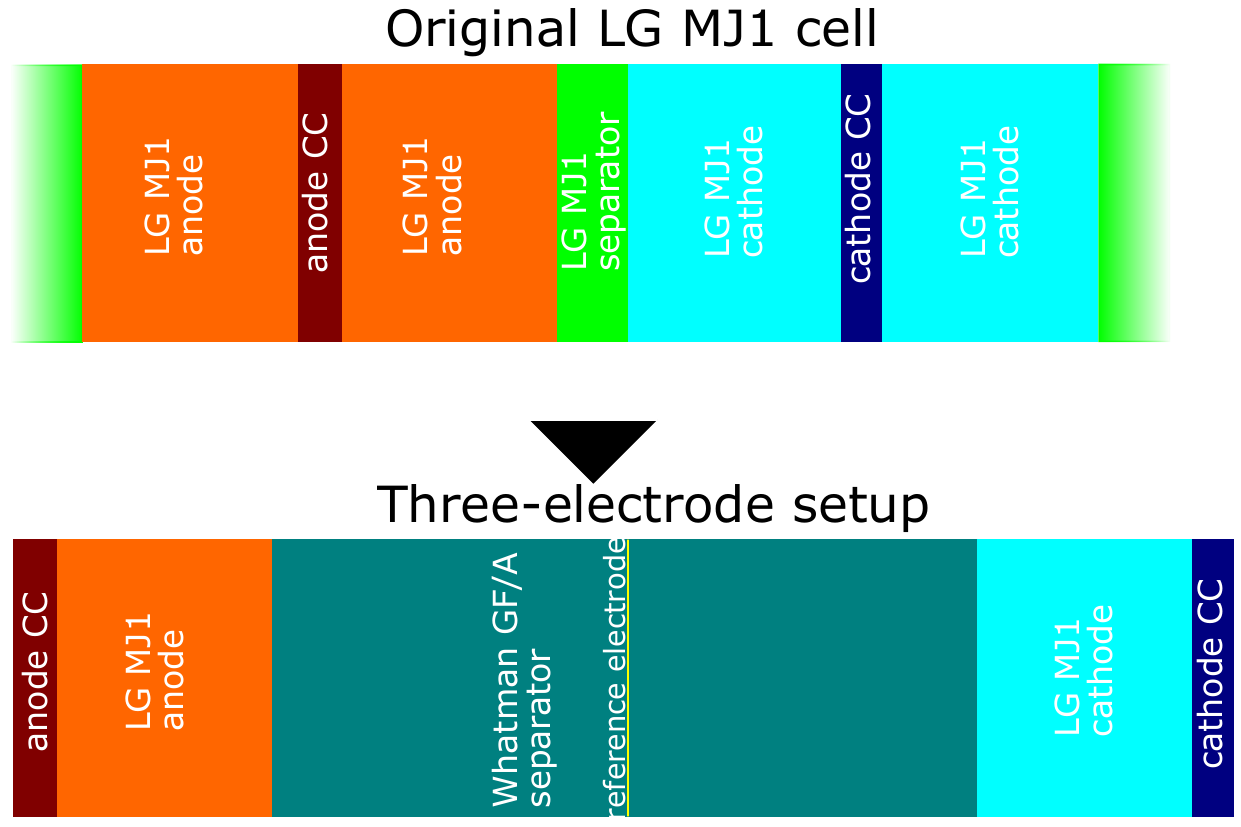}
    \caption{Comparison of the original cell and what was used for the GITT measurements.}
    \label{fig:commercial_to_lab}
\end{center}\end{figure}

\subsection{Microscopy}

\begin{figure}[!ht]\begin{center}
    \includegraphics[width=0.66\textwidth]{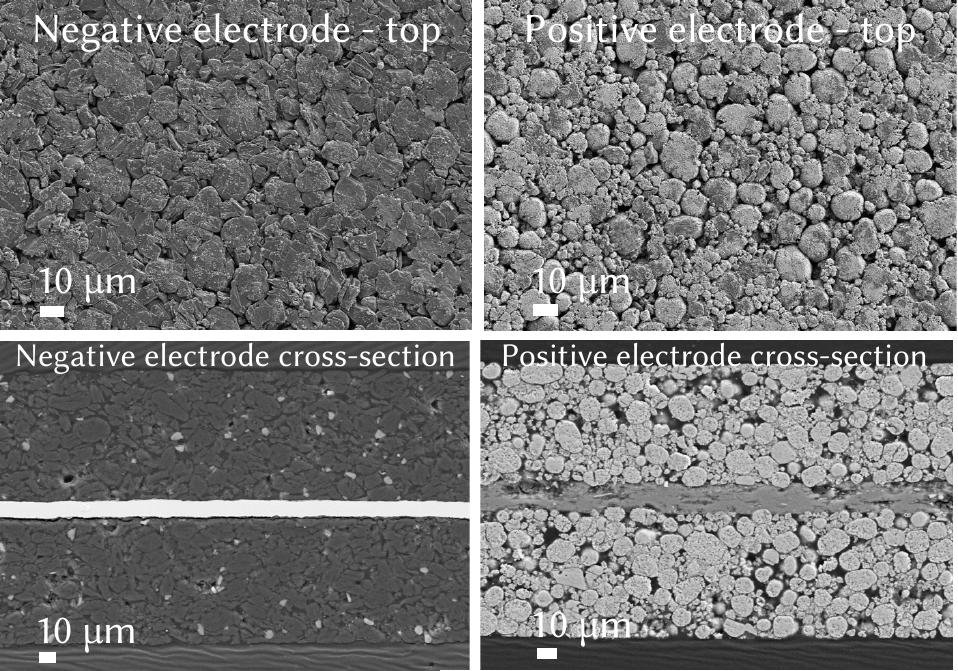}
    \caption{Top (top row) and cross-sectional (bottom row) SEM images of negative electrode (left column) and positive electrode (right column) as extracted from the cell.}
    \label{fig:SEM}
\end{center}\end{figure}

\newpage

\subsection{GITT method elaboration} \label{subsec:GITT}

The original 1977 GITT paper \cite{Weppner1977} derives the diffusion coefficient at flat electrodes as a simple function of concentration overpotential \(\Delta U_t\) over static voltage step \(\Delta U_s\), see \prettyref{eq:GITT}.
\begin{equation}
    D_s \approx \frac{4}{\pi} \frac{\Delta t}{\Delta SOC^2} \left( \frac{I}{z F A c_{s,max}} \frac{\Delta U_s}{\Delta U_t} \right)^2. \label{eq:GITT}
\end{equation}

Here, \(\Delta t\) refers to the duration of the applied constant-current \(I\)\ pulse, \(\Delta SOC\) is the change in state-of-charge during that pulse, \(z=1\) is the charge number of the Lithium ion, \(F\) is the Faraday constant, \(A\) is the cross-section area of the electrode concerning the direction between the current collectors, and \(c_{s,max}\) is the maximum concentration of Lithium atoms in the electrode active material. \par
The 1977 GITT paper also contains a more accurate derivative formulation of GITT, which is the basis for the \(\Delta\)-formulation:
\begin{equation}
    D_s \approx \frac{4}{\pi} \left( \frac{I}{z F A c_{s,max}} \frac{\frac{d OCV}{d SOC}}{\frac{d (U(t) - OCV(SOC))}{d \sqrt{t}}} \right)^2. \label{eq:DerivativeGITT}
\end{equation}

While the original formulas were derived for flat electrodes, they still hold for spherical electrode particles, albeit with a shorter applicable timespan in the measurement. \cite{Chien2021}
Precision may be further enhanced by modifying the square-root transformation such that it accounts for the diffusion relaxation behaviour that started at the beginning of a pulse with duration \(\tau\), which overlaps with the diffusion relaxation behaviour starting at the beginning of the following rest. \cite{Kang2021}
\begin{equation}
    D_s \approx \frac{4}{\pi} \left( \frac{I}{z F A c_{s,max}} \frac{\frac{d OCV}{d SOC}}{\frac{d (U(t) - OCV(SOC))}{d (\sqrt{t + \tau} - \sqrt{t})}} \right)^2.
\end{equation}

\newpage

\section{Workflow example} \label{sec:raw_to_interoperable}

The software implementation of the interpretation step in Python is part of the GitHub repository mentioned in the main document. For information about it, we direct you to its own documentation there. \par
\begin{figure}[ht]\begin{center}
    \includegraphics[width=\textwidth]{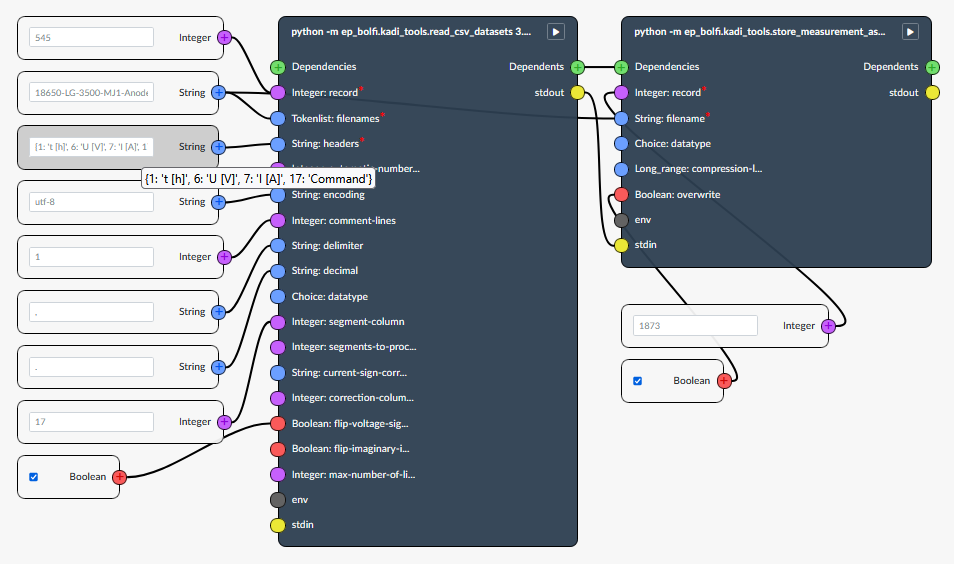}
    \caption{The Kadi4Mat workflow file for raw measurement treatment of the GITT experiment in lithiation direction.}
    \label{fig:raw_to_interoperable}
\end{center}\end{figure}

Here, we showcase the general usage on the script for interpretation of raw measurement files. The visualization of one corresponding Kadi4Mat workflow file is in \prettyref{fig:raw_to_interoperable}. Note how the version of the script used is automatically documented alongside its invocation.

The cycler logged its state in column 17 in the lithiation direction, while it used column 18 in the delithiation direction. The columns of interest are time, current, voltage, and cycler state.
We flip the voltages such that they are positive
We adjust current signs by state column such that they are positive when the cell gets charged/lithiated and negative when it gets discharged/delithiated, relating to the working electrode.

\newpage

\section{GITT analysis results for the NMC electrode} \label{sec:gitt_results}

\begin{figure}[ht]\begin{center}
    \includegraphics[width=0.56\columnwidth]{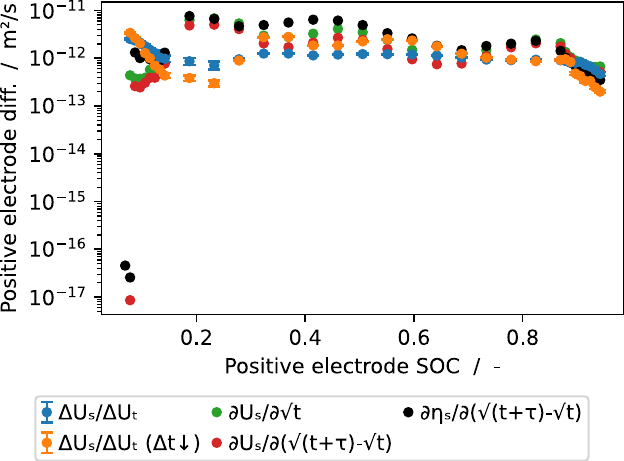}
    \caption{Results for the diffusivity of NMC in delithiation direction, as gleaned with direct calculations from the same GITT data each time.}
    \label{fig:gitt_diffusivity_direct}
\end{center}\end{figure}

\begin{figure}[ht]\begin{center}
    \includegraphics[width=0.56\columnwidth]{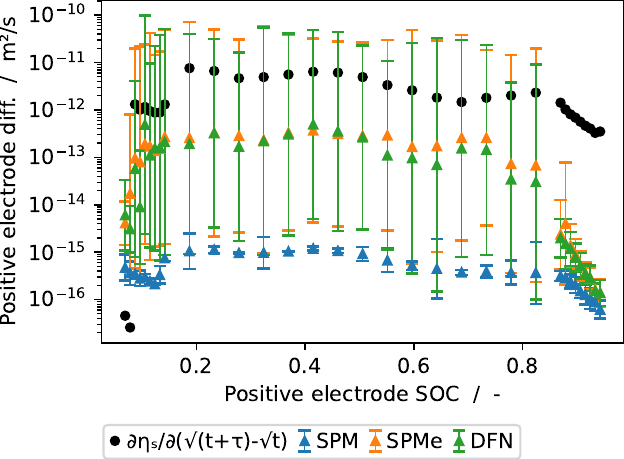}
    \caption{Results for the diffusivity of NMC in delithiation direction, as gleaned from fitting electrochemical models to GITT data, plus the best direct approach.}
    \label{fig:gitt_diffusivity_inverse}
\end{center}\end{figure}

\begin{figure}[ht]\begin{center}
    \includegraphics[width=0.56\columnwidth]{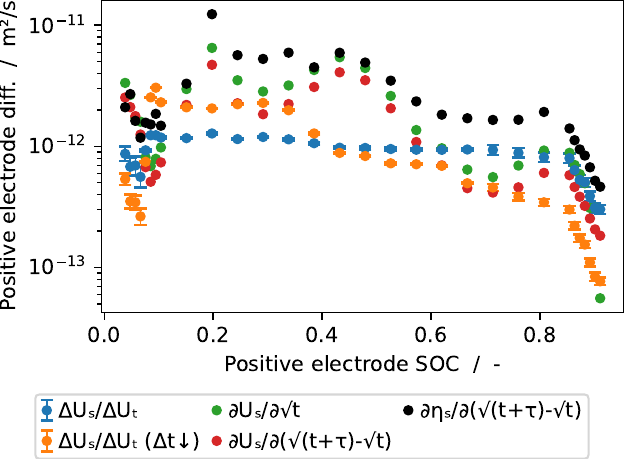}
    \caption{Results for the diffusivity of NMC in lithiation direction, as gleaned with direct calculations from the same GITT data each time.}
    \label{fig:gitt_diffusivity_direct_lithiation}
\end{center}\end{figure}

\begin{figure}[ht]\begin{center}
    \includegraphics[width=0.56\columnwidth]{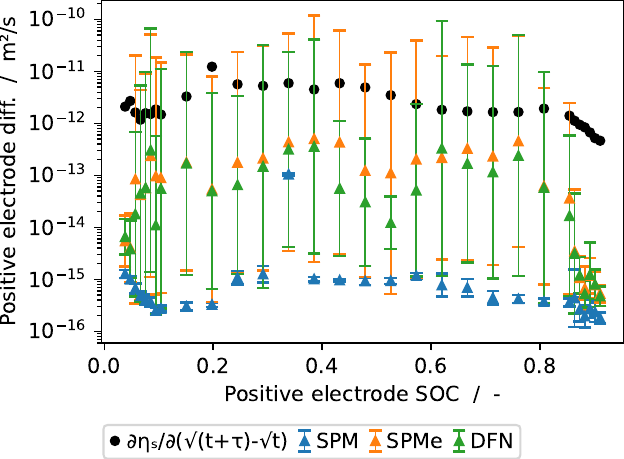}
    \caption{Results for the diffusivity of NMC in lithiation direction, as gleaned from fitting electrochemical models to GITT data, plus the best direct approach.}
    \label{fig:gitt_diffusivity_inverse_lithiation}
\end{center}\end{figure}

\newpage

\bibliographystyle{plain}
\bibliography{library}